\documentstyle[aaspp4,tighten,flushrt]{article}
\newcommand{\beq}{\begin{equation}}
\newcommand{\eeq}{\end{equation}}
\newcommand{\beqa}{\begin{eqnarray}}
\newcommand{\eeqa}{\end{eqnarray}}

\def\IRAS{{\it IRAS} }

\def\lexp{\mathop{{}{\langle}{}}}
\def\rexp{\mathop{{}{\rangle}{}}}
\newcommand{\rexpc}{\mathop{\rangle_c}}

\def\d{\delta}

\def\ds{\delta_s}
\def\dt{\tilde \delta}
\def\dD{[\delta_{\rm D}]}
\def\del{\nabla}

\font\BFd=cmmib10
\font\BFt=cmmib10
\font\BFs=cmmib10 scaled 700
\font\BFss=cmmib10 scaled 500

\def\bbox#1{%
\relax\ifmmode
\mathchoice
{{\hbox{\BFd #1}}}
{{\hbox{\BFt #1}}}
{{\hbox{\BFs #1}}}
{{\hbox{\BFss #1}}}
\else \mbox{#1} \fi }

\def\k{{\bbox{k}}}
\def\x{{\bbox{x}}}
\def\r{{\bbox{r}}}
\def\s{{\bbox{s}}}

\def\q{{\bbox{q}}}
\def\uz{{\hbox{$u_z$}}}
\def\u{{\bbox{u}}}
\def\v{{\bbox{v}}}

\begin{document}

\title{The Bispectrum: From Theory to Observations}

\author{Rom\'an Scoccimarro}

\affil{Institute for Advanced Study, Einstein Drive, Princeton NJ
08540\footnote{current address. email: scoccima@ias.edu}}
\affil{CITA, McLennan Physical Labs, 60 St George Street, 
Toronto ON M5S 3H8, Canada}

\begin{abstract}

The bispectrum is the lowest-order statistic sensitive to the shape of
structures generated by gravitational instability and is a potentially
powerful probe of galaxy biasing and the Gaussianity of primordial
fluctuations.  Although the evolution of the bispectrum is well
understood theoretically from non-linear perturbation theory and
numerical simulations, applications to galaxy surveys require a number
of issues to be addressed.  In this paper we consider the effect on
the bispectrum of stochastic non-linear biasing, radial redshift
distortions, non-Gaussian initial conditions, survey geometry and
sampling.

We find that: 1) bias stochasticity does not affect the use of the
bispectrum to recover the mean biasing relation between galaxies and
mass, at least for models in which the scatter is uncorrelated at
large scales. 2) radial redshift distortions do not change
significantly the monopole power spectrum and bispectrum compared to
their plane-parallel values.  3) survey geometry leads to
finite-volume effects which must be taken into account in current
surveys before comparison with theoretical predictions can be made.
4) sparse sampling and survey geometry correlate different triangles
leading to a breakdown of the Gaussian likelihood approximation.

We develop a likelihood analysis using bispectrum eigenmodes, 
calculated by Monte Carlo realizations of mock surveys generated with 
second-order Lagrangian perturbation theory and checked against N-body 
simulations.  In a companion paper we apply these results to the 
analysis of the bispectrum of \IRAS galaxies.

\end{abstract}

\subjectheadings{large-scale structure of universe}

\clearpage 

\section{Introduction}

It is widely accepted that the large-scale structure of the universe
routinely observed in galaxy surveys is the result of gravitational
instability that operates on (initially small) fluctuations in the
density distribution. Essentially, initially dense (underdense)
regions contract (expand) as a consequence of the attractive nature of
gravity. As a result, most of the volume of the universe is in
underdense regions whereas the rest contains dense structures such as
clusters, galaxies, etc. This asymmetry between underdense and
overdense regions implies non-Gaussianity in the galaxy
distribution. Understanding the dynamics of gravitational instability
thus gives us a powerful tool that can be used to extract useful
information from the non-Gaussian clustering pattern of galaxies
(Peebles 1980; Fry 1984; Goroff et al. 1986; Bernardeau 1992; Bouchet
et al. 1995; Scoccimarro \& Frieman 1996).

A quantitative description of galaxy clustering is obtained by
applying statistical methods, in particular, multi-point correlation
functions (Peebles 1980). In a Gaussian field, the statistical
properties are fully characterized by its two-point correlation
function or power spectrum, all higher-order (connected) correlation
functions being zero. Information on non-Gaussianity is thus
accessible by calculating higher-order correlations functions. An
alternative to multi-point correlation functions is to consider
higher-order moments (i.e. smoothed multi-point correlation functions
collapsed at a single point); these characterize the one-point
distribution of the galaxy distribution and can be measured with
larger signal to noise than the corresponding multi-point
functions. However, the latter contain more information, in
particular, being evaluated at three or more points they are sensitive
to the shapes of large-scale structures, which is the key to
disentangle degeneracies present otherwise.

The bispectrum, the three-point correlation function in Fourier space,
is the lowest order statistic sensitive to the shape of structures
generated by gravitational instability. It can be used to probe the
nature of primordial fluctuations (Gaussian versus non-Gaussian),
galaxy biasing (i.e.  the relation between the galaxy distribution and
the underlying dark matter distribution), and help to break the
degeneracy between linear bias and the matter density parameter
$\Omega_m$ present in power spectrum measurements in galaxy redshift
surveys (Fry 1994; Hivon et al. 1995; Matarrese, Verde \& Heavens
1997; Scoccimarro et al. 1998).

Pioneering work on three-point statistics from the Zwicky (Peebles \&
Groth 1975) and Lick (Groth \& Peebles 1977; Fry \& Seldner 1982)
angular catalogs showed that the bispectrum, $B(k_1,k_2,k_3)$, scaled
in terms of the power spectrum, $P(k)$, as $B(k) \propto P(k)^2$. This
motivatied the so-called hierarchical model and was also in agreement
with the scaling at small scales inferred from the BBGKY equations of
motion assuming stable clustering and self-similarity (Davis \&
Peebles 1977; Peebles 1980). At these small scales deep into the
non-linear regime, however, limited understanding of the non-linear
physics including the possibility of complicated galaxy biasing
prevents using higher-order statistics to extract quantitative
knowledge.

It was not until somewhat later, with the systematic development of
non-linear perturbation theory (hereafter PT; Fry 1984), that it was
possible to derive quantitatively how non-Gaussianity is generated by
gravity. In particular, the scaling $B(k) \propto P(k)^{2}$ (and in
general, the $n$-point function $\xi_n$ scales as $\xi_n \propto
\xi_2^{n-1}$) is recovered at large scales, being simply a result of
quadratic non-linearities in the equations of motion of gravitational
instability. An important signature predicted by PT is that the
bispectrum at large scales should be a rather strong function of
triangle configuration, with the reduced bispectrum $Q_{123}$ 

\beq 
Q_{123} = \frac{B_{123}}{P_{1}P_{2}+ P_{2}P_{3}+ P_{3}P_{1}}
\label{q},
\eeq
\noindent where $B_{123}\equiv B(k_1,k_2,k_3)$ and $P_i\equiv P(k_i)$,
showing higher amplitude for colinear configurations than equilateral
configurations, corresponding to the fact that gravity generates
anisotropic large-scale structures.  These results have
been confirmed since then by N-body simulations (Fry, Melott \&
Shandarin 1993; Scoccimarro et al. 1998).

These predictions hold for the density field, in order to compare with
observations one must necessarily deal with the possibility of galaxy
biasing. At scales larger than those relevant to galaxy formation, one
can make the simplifying assumption that biasing is a local function
of the underlying {\em smoothed} density field, and thus expandable
perturbatively at large scales as (Fry \& Gazta\~naga 1993; see also
Coles 1993)

\beq
\dt_g(\x) = \sum_{n=0} b_n\ \dt^n(\x),
\label{locbias}
\eeq
\noindent where $b_n$ are the bias parameters, and a tilde denotes the
smoothing operation.  Thus the galaxy bispectrum is calculable in
terms of the density bispectrum at leading order in PT

\beq 
Q_g = {Q \over b_1} + {b_2 \over b^2_1}
\label{Qg}
\eeq
\noindent in terms of linear ($b_1$) and non-linear ($b_2$) biasing
parameters. Basically, a large linear bias decreases the dependence of
the bispectrum on triangle, whereas antibias ($b_1<1$) enhances the
configuration dependence of the galaxy bispectrum relative to that of
the density.  Since the density bispectrum depends to a good
approximation only on spectral index, comparison of the density and
galaxy bispectrum can give constraints on the biasing parameters
independent of $\Omega_m$ (Fry 1994). Note that $Q_g$ and $Q$ are {\em
functions} that depend on triangle configuration; this is what allows
to break the degeneracy between $b_1$ and $b_2$ present in analogous
measurements of one-point moments using the skewness. Note that
Eq.~(\ref{Qg}) assumes Gaussian initial conditions, otherwise there
are additional contributions (Scoccimarro 2000).

A first attempt to use this procedure in the Lick catalog data,
concluded that the lack of configuration dependence on the observed
bispectrum required a large bias, $b_{1} \approx 3$, assuming the
absence of systematic errors (Fry 1994). However, other effects such
as projection (Fry \& Thomas 1999; Frieman \& Gazta\~naga 1999;
Buchalter, Kamionkowski \& Jaffe 2000) and most importantly non-linear
evolution (Scoccimarro et al.  1998) also wash out the configuration
dependence of the bispectrum, so application of this idea to
observational data had to await until the availability of larger
galaxy surveys. Recent determination of the angular three-point
correlation function from the APM survey (Frieman \& Gazta\~naga 1999)
shows a configuration dependence consistent with APM galaxies being
unbiased ($b_1\approx 1$, $b_2\approx 0$), but with large error
bars. This result is however consistent with the determination of
one-point higher-order moments in the APM (Gazta\~naga 1994), which
show remarkable agreement with the predictions of PT from Gaussian
initial conditions and no bias (Juszkiewicz, Bouchet \& Colombi 1993;
Bernardeau 1994a,1994b).

Redshift surveys map the three-dimensional distribution of galaxies in
a large volume, and are thus ideally suited to use the bispectrum as a
probe of galaxy biasing and non-Gaussian initial conditions. In this
case, the observed distribution is actually distorted by peculiar
velocities along the line of sight which contribute to
redshifts. However, these redshift-space distortions are now well
understood, at least in the plane-parallel approximation (Hivon et al.
1995, Verde et al. 1998; Scoccimarro, Couchman \& Frieman
1999). Determination of three-point statistics from redshift surveys
has been carried out for the CfA survey and a sample of redshifts in
the Pisces-Perseus supercluster (Baumgart \& Fry 1991) and the LCRS
survey (Jing \& B\"orner 1998). In both cases, however, there is a
rather limited range of scales in the weakly non-linear regime to
obtain quantitative constraints on biasing and non-Gaussian initial
conditions. On the other hand, the next generation of large redshift
surveys such as 2dF and SDSS will provide excellent conditions for the
determination of the bispectrum at large scales.

There are however a number of issues which are potentially important
for current and future surveys and must be addressed to reliably use
the bispectrum to extract useful cosmological information.  In this
paper we consider effects that can potentially influence theoretical
predictions of the bispectrum in redshift surveys: stochastic biasing,
radial nature of redshift distortions, finite survey volume and sparse
sampling.  The main result of this study is a likelihood analysis that
allows extracting accurate constraints on galaxy biasing and
non-Gaussian initial conditions.  In this work, we consider \IRAS mock
surveys, which map a large enough volume in the weakly non-linear
regime. Application of our results to the actual data is considered in
a companion paper (Scoccimarro et al. 2000).

This paper is organized as follows. In Section~2 we briefly review the
behavior of the bispectrum at large scales from PT. In Section~3 we
discuss the effects of galaxy biasing, including stochasticity. In
Section~4 we review second-order Lagrangian PT, and discuss its regime
of validity by comparing to N-body simulations. Section~5 presents
results on the effects of redshift distortions in the plane-parallel
and radial case. Section~6 discusses the generation of mock catalogs
and the optimal weighting for higher-order statistics. In Section~7 we
discuss the effects of finite volume of the survey and sampling on the
bispectrum. Section~8 develops the likelihood analysis using
bispectrum eigenmodes.  Finally, Section~9 offers a summary of the
results and our conclusions.

\section{The Bispectrum induced by Gravity}
\label{bgrav}

If primordial density perturbations are Gaussian, they are fully
characterized by their power spectrum

\beq
\lexp \d(\k_1) \d(\k_2) \rexp =  \dD_{12}\ P(k_1),
\eeq

\noindent where $\dD_{1 \ldots N} \equiv (2\pi)^{3} \d_D(\k_1 +
\ldots + \k_N)$, with $\d_D(\x)$ the Dirac delta distribution. In this
section we assume that the density field obeys statistical isotropy
and homogeneity; these assumptions break down in presence of
redshift-distortions, as will be discussed below. Even though the
initial conditions are Gaussian, gravitational clustering is a
non-linear process that induces non-Gaussianity through mode-mode
coupling. In particular, the three-point cumulant in Fourier space,
the bispectrum

\begin{equation}
\lexp \d(\k_1) \d(\k_2) \d(\k_3) \rexpc =  \dD_{123} \
B_{123}
\label{bispectrum},
\end{equation} 

\noindent becomes non-zero, the leading order contribution arising
from second-order PT (\cite{Fry84b}). As discussed in the
Introduction, since $B(k) \propto P(k)^2$, it is convenient to work
with the reduced bispectrum $Q_{123}$ (\cite{FrySeldner82}) defined as
in Eq.~(\ref{q}), which has the desirable property that it is time and
approximately scale independent to lowest order (tree-level) in
non-linear PT, i.e.

\begin{equation}
B_{123}=2 F_2(\k_1,\k_2)\ P_1 P_2 + {\rm cyc.}
\label{btree},
\end{equation} 
 
\noindent with 

\beq 
\d_2(\k)= \int d^3q\ F_2(\q,\k-\q)\ \d_1(\q) \d_1(\k-\q),
\label{d2} 
\eeq
\beq 
F_2(\k_1,\k_2)= \frac{5}{7}+ \frac{\k_1 \cdot \k_2}{2 k_1 k_2}
\left(\frac{k_1}{k_2}+\frac{k_2}{k_1}\right) + \frac{2}{7}
\left( \frac{\k_1 \cdot \k_2}{k_1 k_2} \right)^2, 
\label{F2}
\eeq

\noindent where $\d_2(\k)$ denotes the solution to the equations of
motion of gravitational instability to second order in PT
(\cite{Fry84b,GGRW86}). The result in Eq.~(\ref{F2}) is for an
Einstein-de Sitter model, but an important property of the kernel
$F_2$ is its very weak dependence on density parameter $\Omega_m$. A
good approximation is to replace $5/7 \rightarrow (1+\kappa)/2$ and
$2/7 \rightarrow (1-\kappa)/2$, where for an open Universe $\kappa
\approx \frac{3}{7} \Omega_m^{-2/63}$, and for a flat Universe with
cosmological constant, $\kappa \approx \frac{3}{7} \Omega_m^{-1/143}$
(\cite{BJCP92,BCHJ95}). For scale-free initial conditions ($P(k)
\propto k^n$), the density field reduced bispectrum at large scales
thus becomes a function of only the spectral index $n$, and the shape
of the triangle through e.g. the ratio between two sides $k_1/k_2$ and
the angle between them.

So far we have discussed results for Gaussian initial conditions. When
this assumption is relaxed, higher-order correlation functions are
non-zero from the begining and the evolution of three-point statistics
beyond linear PT is non-trivial (\cite{FrSc94}; Scoccimarro 2000). To
illustrate this consider the tree-level evolution of the bispectrum
from arbitrary non-Gaussian initial conditions

\begin{eqnarray}
B_{123}^{(0)} &=& B_{123}^I + B_{123}^G + \int d^3q\
F_2(\k_1+\k_2-\q,\q)\ T_4^I(\k_1,\k_2,\k_1+\k_2-\q,\q),
\label{BispNG}
\end{eqnarray}

\noindent where $B_{123}^I$ denotes the contribution of the initial
bispectrum, scaled to the present time using linear PT, $B_{123}^G$
represents the usual gravitationally induced bispectrum, and the last
term represents the contribution coming from the initial trispectrum
linearly evolved to the present, $T_4^I$ given by

\begin{equation}
\left<\d^I(\k_1)\d^I(\k_2)\d^I(\k_3)\d^I(\k_4) \right>_c \equiv
\dD_{1234} \ T_4^I
\label{trisp}.
\end{equation}
 
\noindent In this work we will consider one particular model of
non-Gaussian initial conditions, the $\chi^2$ model (e.g. Kofman et
al. 1989; Antoniadis et al. 1997, Linde \& Muhanov 1997; Peebles
1997).

\section{Galaxy Biasing: Effects of Stochasticity}
\label{bias}

As discussed in the Introduction, when interpreting clustering
statistics measured in galaxy surveys, one must necessarily deal with
the issue of galaxy biasing. Equation~(\ref{locbias}) assumes not only
that the bias is local (which seems a reasonable assumption at large
scales) but also deterministic; that is, the galaxy distribution is
completely determined by the underlying mass distribution. In
practice, however, it is likely that galaxy formation depends on other
variables besides the density field, and that consequently the
relation between $\dt_g(\x)$ and $\dt(\x)$ is not deterministic but
rather stochastic, 

\beq
\dt_g(\x) = \sum_{n=0} b_n\ \dt^n(\x) + \varepsilon_\d(\x),
\label{locbias2}
\eeq

\noindent where the random field $\varepsilon_\d(\x)$ denotes the
scatter in the biasing relation {\em at a given} $\d$ due to the fact
that $\dt(\x)$ does not completely determine $\dt_g(\x)$. Recent work
has focussed on the effects of stochasticity on one-point statistics,
the two-point correlation function and power spectrum (Scherrer \&
Weinberg 1998; Dekel \& Lahav 1999). Clearly, if the field
$\varepsilon(\x)$ is arbitrary, the effects on clustering statistics
could be rather strong. However, at large enough scales, we expect the
field $\varepsilon_\d(\x)$ to be weakly correlated; that is, at large
smoothing scales the scatter about the mean bias relation should be
local. This means that $\lexp \varepsilon_\d(\x_1)
\varepsilon_\d(\x_2) \rexp \approx \sigma_\varepsilon^2
\Theta(a-|\x_1-\x_2|)$, where $\Theta(x)=1$ if $x>0$ and zero
otherwise, $a$ is the correlation ``range'' of the scatter and
$\sigma_\varepsilon^2$ its strength. In this case, the power spectrum
reads

\beq
P_g(k) = b_1^2 P(k) + \sigma_\varepsilon^2 V_a W_{\rm TH}(ka) ,
\label{Pg}
\eeq

\noindent where $V_a=4\pi a^3/3$, and $W_{\rm TH}(x)=3(\sin x-x \cos
x)/x^3$ is the Fourier transform of the top-hat window. Here we used
that, by definition, $\lexp \varepsilon \d^n \rexp =0$. So, at large
scales, $ka \ll 1$, stochastic bias leads to a constant offset to the
power spectrum (Scherrer \& Weinberg 1998; Dekel \& Lahav 1999),
similarly to Poisson fluctuations due to shot noise. However, in
principle, the scatter at a given $\d$ could be non-Poissonian
(e.g. see Sheth \& Lemson 1999). As long as the second term in
Eq.~(\ref{Pg}) is small enough, the effect of local scatter should not
very important. In this paper we do not explore general models of
stochastic bias (see e.g. Scherrer \& Weinberg 1998; Dekel \& Lahav
1999; Blanton et al. 1999; Matsubara 1999), but rather test whether
stochasticity affects the determination of bias parameters from
bispectrum measurements that assume Eq.~(\ref{Qg}) in simple models
where the scatter is local but not negligible when the ``galaxy''
density field is smoothed with a Gaussian filter of radius $R_s
\approx 10$ Mpc/h. A similar approach in the context of higher-order
one-point moments is given by Narayanan, Berlind, \& Weinberg (2000),
who also consider the effect of some non-local models of galaxy
biasing.

We follow Cole et al. (1998) and generate non-linearly stochastic
biased density fields by selecting dark matter particles to be
``galaxies'' with a probability function (see also Mann, Peacock, \&
Heavens 1998)

\beqa 
P(\nu) &\propto& \exp( a \nu + b \nu^{3/2} )\ \ \ \ \ 
(\nu\geq 0) \label{probaa} \\ P(\nu) &\propto& \exp( a \nu )\
\ \ \ \ (\nu< 0), 
\label{proba} 
\eeqa

\noindent where $\nu(\x) \equiv \dt(\x)/\sigma$ is the local
normalized smoothed density field. The model has two free parameters
$a$ and $b$ that can be adjusted to generate stochastic biased fields
that are preferentially weighted towards low or high density
regions. The mass density field is generated using 2LPT (see
description in Section 4.2 below) in a 300 Mpc/h box, with a $128^3$
grid. The smoothed density field in Eqs.(\ref{probaa}-ref{proba}) then
corresponds to about 3 Mpc/h smoothing. For a given particle, we
choose the nearest grid point to evaluate $\nu(\x)$, thus the biasing
scheme is effectively non-local at scales of about 3 Mpc/h.

Figure~\ref{fig_bias} shows four examples of such a procedure. The top
panels correspond to bias towards high density regions, with
$b_1=b_2=1.54$ (left panel) and $b_1=1.45$, $b_2=1.06$ (right panel),
whereas the bottom panels show examples of bias that preferentially
select low density regions with $b_1=0.95$, $b_2=-0.06$ (left panel)
and $b_1=0.77$, $b_2=-0.14$ (right panel). All except the bottom right
panel are from $\Lambda$CDM realizations ($\Omega_m=0.3$,
$\Omega_\Lambda=0.7$) with $\sigma_8=0.70$ for the dark matter. The
bottom right panel corresponds to a $\tau$CDM realization ($\Omega_m=1$)
with same normalization and power spectrum shape. The two right panels
have roughly the same redshift distortions parameter $\beta \equiv
\Omega_m^0.6/b_1 \approx 0.65$, and thus will be nearly equal in terms
of their redshift-space power spectrum. These bias parameter values in
Figure~\ref{fig_bias} have been obtained by measuring the bispectrum
for the ``galaxy'' distribution and comparing to the mass bispectrum
using Eq.~(\ref{Qg}), and are plotted in each panel as the continuous
solid line. The actual mean of the relation and its 90\% confidence
region for smoothed fields with Gaussian filter radius $R_s=10$Mpc/h
show that even when the $\d-\d_g$ relation has considerable scatter,
the bispectrum recovers the mean of the biasing relation quite
accurately. Below, we shall discuss whether this result is also valid
when dealing with noisy data such as in a galaxy survey, including the
effects of survey geometry and sampling.

\section{Numerical Implementation of PT: Comparison with N-body Simulations}

\subsection{The Need for Numerical PT}

In order to understand how geometry and sampling affects statistics
such as the bispectrum, we need to simulate galaxy catalogs and
measure their statistical properties to compare with the ideal case
known from analytical PT calculations and N-body simulations. As we
shall see, it will be become very valuable to make a large number (a
few hundred to a thousand) of mock catalogs for different cosmological
models and survey selection functions. As a result, it would be very
expensive to proceed with N-body simulations, even with PM
simulations. Since we are interested in the large-scale properties of
clustering, where PT applies, it is natural to resort to a numerical
implementation of PT.

Lagrangian PT is the natural choice. In this formulation, particles
are displaced from their initial positions by a displacement field
that is found order by order perturbatively from the equations of
motion. The linear solution is the well-known Zel'dovich (1970)
approximation (ZA), and higher-order corrections have been well
studied in the literature (\cite{MABPR91,BMW94,BCHJ95}). Although the
ZA is the least expensive of them, it does not reproduce very
accurately the statistical properties of clustering even at large
scales (\cite{GrWi87,JBC93,Bernardeau94c,CaMo94,JWACB95}). In
particular, we are interested in studying the bispectrum, and the ZA
overestimates its configuration dependence and underestimates its
overall amplitude. The situation becomes even worse as higher-order
statistics are considered. To illustrate this, Table~\ref{ZASp} shows
the ratio of the $S_p$ parameters in the ZA to their exact values at
large scales as a function of spectral index $n$. The $S_p$ parameters
characterize the one-point PDF of the smoothed density field at scale
$R$ (\cite{GGRW86})

\beq
S_p \equiv \frac{\lexp \dt^p \rexpc}{\lexp \dt^2 \rexp^{p-1}},
\label{Sp}
\eeq

\noindent and Table~\ref{ZASp} shows $r_p \equiv S_p^{\rm ZA}/ S_p$,
where $S_p$ is the exact value calculated in full PT
(\cite{JBC93,Bernardeau94a,Bernardeau94c}). We see that for $p=3$ the
skewness is poorly reproduced as the spectral index becomes positive,
in the relevant range for CDM models at large scales. As a result,
using the ZA to generate mock catalogs to study the statistical
properties of the bispectrum could result in serious quantitative
error.

The next level of accuracy is second-order Lagrangian PT (2LPT), which
reproduces the three-point statistics (skewness and bispectrum)
exactly, and gives reasonable values for higher-order statistics as
well, in remarkable improvement over the ZA
(\cite{JWACB95}). Table~\ref{2LPTSp} shows the analogous results for
$r_p$ in 2LPT (\cite{Scoccimarro98}). Given the small additional
computational expense of 2LPT over ZA (see discussion below), it is a
very convenient scheme to implement for mock catalog production. It is
still orders of magnitude faster than a PM simulation. The question
then arises, is it worth to consider going beyond 2LPT? At
third-order, 3LPT recovers four-point statistics exactly (see
Table~\ref{3LPTSp}), but it involves an appreciable increase in
complexity compared to 2LPT, requiring to solve three additional
Poisson's equations (\cite{BMW94}). For this reason, we consider 2LPT
the optimal choice between speed and accuracy. We now briefly discuss
2LPT and its implementation.

\subsection{Second-Order Lagrangian PT (2LPT)}

In Lagrangian PT, the object of interest is the displacement field
${\bf \Psi}(\q)$ which maps the initial particle positions $\q$ into
the final Eulerian particle positions $\x$,

\beq
\x = \q + {\bf \Psi}(\q).
\eeq

\noindent The equation of motion for particle trajectories $\x(\tau)$
is

\beq
\frac{d^2 \x}{d \tau^2} + {\cal H}(\tau) \ \frac{d \x}{d \tau}= -
\del \Phi, 
\eeq

\noindent where $\Phi$ denotes the gravitational potential, and $\del$
the gradient operator in Eulerian coordinates $\x$. Taking the
divergence of this equation we obtain

\beq
J(\q,\tau)\ \del \cdot \Big[ \frac{d^2 \x}{d \tau^2} + {\cal
H}(\tau) \ \frac{d \x}{d \tau} \Big] = \frac{3}{2} \Omega_m {\cal H}^2
(J-1),
\label{leom}
\eeq

\noindent where we have used Poisson equation together with the fact
that $1+\d(\x) =J^{-1}$, and the Jacobian $J(\q,\tau)$ is the
determinant

\beq
J(\q,\tau)  \equiv  {\rm Det}\Big( \d_{ij}+ \Psi_{i,j} \Big),
\label{jacobian}
\eeq

\noindent where $\Psi_{i,j} \equiv \partial\Psi_i /\partial
\q_j$. Equation~(\ref{leom}) can be fully rewritten in terms of
Lagrangian coordinates by using that $\del_i = ( \d_{ij}+
\Psi_{i,j})^{-1} \del_{q_j}$, where $\del_q \equiv \partial /\partial
\q$ denotes the gradient operator in Lagrangian coordinates. The
resulting non-linear equation for ${\bf \Psi}(\q)$ is then solved
perturbatively, expanding about its linear solution, the Zel'dovich
(1970) approximation

\beq
\del_q \cdot {\bf \Psi}^{(1)}= -D_1(\tau) \ \d(\q),
\label{Psi1}
\eeq 

\noindent which incorporates the kinematic aspect of the collapse of
fluid elements in Lagrangian space. Here $\d(\q)$ denotes the
(Gaussian) density field imposed by the initial conditions and
$D_1(\tau)$ is the linear growth factor. The solution to second order
describes the correction to the ZA displacement due to gravitational
tidal effects and reads

\beq
\del_q \cdot {\bf \Psi}^{(2)}= \frac{1}{2} D_2(\tau) \sum_{i \neq j}
(\Psi_{i,i}^{(1)} \Psi_{j,j}^{(1)} - \Psi_{i,j}^{(1)} \Psi_{j,i}^{(1)}),
\label{Psi2}
\eeq

\noindent (e.g., Bouchet et al. 1995) where $D_2(\tau)$ denotes the
second-order growth factor, which for $0.1 \leq \Omega_m \leq 3$
($\Lambda=0$) obeys

\beq
D_2(\tau) \approx -\frac{3}{7} D_1^2(\tau) \ \Omega_m^{-2/63} \approx
-\frac{3}{7} D_1^2(\tau) 
\eeq

\noindent to better than 0.5\% and 7\%, respectively (Bouchet et
al. 1995), whereas for flat models with non-zero cosmological constant
$\Lambda$ we have for $0.01 \leq \Omega_m \leq 1$

\beq
D_2(\tau) \approx -\frac{3}{7} D_1^2(\tau) \ \Omega_m^{-1/143} \approx
-\frac{3}{7} D_1^2(\tau),  
\eeq

\noindent to better than 0.6\% and 2.6\%, respectively (Bouchet et
al. 1995).  Since Lagrangian solutions up to second-order are
curl-free, it is convenient to define Lagrangian potentials
$\phi^{(1)}$ and $\phi^{(2)}$ so that in 2LPT

\beq
\x(\q) = \q -D_1\ \del_q \phi^{(1)} + D_2\ \del_q \phi^{(2)},
\label{dis2}
\eeq

\noindent and the velocity field then reads ($t$ denotes cosmic time)

\beq
{\bf v} \equiv \frac{d \x}{d t} = -D_1\ f_1\ H\ \del_q \phi^{(1)}
+ D_2\ f_2\ H\ \del_q \phi^{(2)},
\label{vel2}
\eeq 

\noindent where $H$ is the Hubble constant, and the logarithmic
derivatives of the growth factors $f_i \equiv (d\ln D_i)/(d \ln a)$
can be approximated for open models with $0.1 \leq \Omega_m \leq 1$ by

\beq
f_1 \approx \Omega_m^{3/5}, \ \ \ \ \ f_2 \approx 2 \ \Omega_m^{4/7},
\eeq 

\noindent to better than 2\% (Peebles 1976) and 5\% (Bouchet et
al. 1995), respectively. For flat models with non-zero cosmological
constant $\Lambda$ we have for $0.01 \leq \Omega_m \leq 1$

\beq
f_1 \approx \Omega_m^{5/9}, \ \ \ \ \ f_2 \approx 2 \ \Omega_m^{6/11},
\eeq 

\noindent to better than 10\% and 12\%, respectively (Bouchet et
al. 1995). The accuracy of these two fits improves significantly for
$\Omega_m \geq 0.1$, in the range relevant for the present purposes.
The time-independent potentials in Eqs.~(\ref{dis2}) and~(\ref{vel2})
obey the following Poisson equations (Buchert et al. 1994)

\label{poisson2lpt}
\beqa
\del_q^2 \phi^{(1)}(\q) &=&  \d(\q), 
\label{phi1} \\ 
\del_q^2 \phi^{(2)}(\q)&=&  \sum_{i>j}
[\phi_{,ii}^{(1)}(\q)\ \phi_{,jj}^{(1)}(\q) - (\phi_{,ij}^{(1)}(\q))^2],
\label{phi2}
\eeqa

\subsection{Regime of Validity}

In order to assess the validity of 2LPT, we have run N-body
simulations using the Hydra adaptive P$^3$M code (\cite{CTP95}). We
have run 4 realizations of $\Lambda$CDM ($\Omega_m=0.3$,
$\Omega_\Lambda=0.7$, $\sigma_8=0.90$) and 2 realizations of SCDM
($\sigma_8=0.61$), using a 300 Mpc/h box, with $128^3$ particles and
softening length $\epsilon=0.25$ Mpc/h. The 2LPT realizations were
made using in most cases $256^3$ particles in a 600 Mpc/h box
(identical results are obtained using 300 Mpc/h boxes).

We now use the results of these simulations to check the regime of
validity of 2LPT. We shall concentrate on power spectrum and
bispectrum statistics, in real and redshift space. When dealing with
Lagrangian PT, a signature of its breakdown is the amount of shell
crossing; the fact that the Jacobian in Eq.~(\ref{jacobian}) becomes
zero at the regions where particles (initially at different positions)
cross each other. This effect is well-known in the ZA, and leads to
the ``thickening of pancakes'', that is, to significant smoothing of
high density regions. A way of characterizing this effect is to recall
that the density field in the ZA is given by

\beq
1+\d(\x) = \frac{1}{J(\q,\tau)}=\frac{1}{\prod_{i=1}^{3}
(1+\lambda_i(\q,\tau))},  
\eeq

\noindent where $\lambda_i$ denotes the eigenvalues of the first-order
deformation tensor $\Psi_{i,j}^{(1)}$ at point $\q$ and time $\tau$
[see Eq.~\ref{jacobian}]. The condition for shell-crossing can then be
explicitly written as $1+\lambda_i(\q,\tau)=0$, for any $i=1,2,3$. At
these points, one expects not only a breakdown of the ZA, but also of
higher-order Lagrangian PT, in particular 2LPT. An alternative
indicator of the breakdown of 2LPT, as in any perturbative approach,
is to compare the magnitude of the second-order to the first-order
displacement field. One expects 2LPT will break down when the
second-order contribution becomes significant compared to the ZA. 

Table~\ref{sctable} shows these indicators in 2LPT realizations as
described above for $\Lambda$CDM with normalization $\sigma_8=0.7$. In
order to investigate the dependence of the breakdown of 2LPT with the
amount of small-scale power, we include a cutoff in the linear power
spectrum so that $P_{lin}(k)=0$ for $k>k_c$. As more high-frequency
waves are included in the representation of the linear density field,
one expects 2LPT to break down at larger scales. On the other hand,
not enough waves lead to lack of mode-coupling, and therefore
inaccurate representation of the density field one is trying to
simulate. Thus there is an ``optimum'' cutoff; here we considered
models with $k_c=0.3,0.4,0.5,\infty$ h/Mpc.

In Table~\ref{sctable} we show much shell crossing is developed as a
function of $k_c$, in terms of the quantity $x_{sc} \equiv 100 \times
N_{sc}/N_{grid}^3$, where $N_{sc}$ is the number of grid points that
obeyed the shell-crossing condition ($|\lambda_i| \geq 1$ for at least
one $i=1,2,3$), and $N_{grid}$ is the total number of grid points
(where particles are initially located, $N_{par}=
N_{grid}^3$). Obviously, $x_{sc}$ is an increasing function of
$k_c$. The remaining column in Table~\ref{sctable} shows the ratio of
mean second to first-order displacement field magnitudes. Again, this
ratio increases with $k_c$, as expected. We see that even though this
ratio is always less than unity, significant shell crossing develops
(e.g. for $k_c>0.5$), so breakdown of 2LPT is nonetheless expected to
happen. In the following we use 2LPT realizations with $k_c=0.5$
h/Mpc, which we found reproduces clustering statistics in redshift
space very well.

In Figure~\ref{fig_sc} we show a comparison of 2LPT (solid lines) with
N-body simulations (symbols) in redshift space. The top panel shows
the power spectrum (in terms of $\Delta(k)=4\pi k^3 P(k)$) as a
function of scale, the dotted line corresponds to linear PT,
Eq.~\ref{delta_sl} below (Kaiser 1987). At large scales, $k<0.2$
h/Mpc, 2LPT reproduces the suppresion of power compared to linear PT;
at smaller scales the supression is overestimated, as we approach the
cutoff scale of the power spectrum and also shell crossing starts
playing a role. The bottom panel shows a similar comparison for the
equilateral bispectrum, the dotted line now corresponds to the
redshift space prediction at large scales from tree-level PT
(Scoccimarro et al. 1999). We see that although tree-level PT breaks
down at about $k \approx 0.2$ h/Mpc, 2LPT continues to hold at smaller
scales, $k \leq 0.4$ h/Mpc. Figure~\ref{fig_bktheta} shows a
comparison for the redshift space bispectrum for configurations in
which $k_2=2k_1=0.21$ h/Mpc as a function of the angle $\theta$
between $\k_1$ and $\k_2$. The dotted line shows the prediction of
tree-level PT (Hivon et al. 1995). As noted before (Scoccimarro et al
1999), the predictions of tree-level PT in redshift space break down
at relatively large scales, due to the non-perturbative nature of the
redshift-space mapping. Fortunately, since 2LPT performs the mapping
exactly, the agreement with the numerical simulations is very good
down to scales $k \approx 0.4$ h/Mpc.  For our purposes, we are
interested in studying statistics at large scales, $k \leq 0.2$ h/Mpc
(with $k_{nl} > 0.2$ h/Mpc), so the use of 2LPT is well justified.

\section{Redshift Distortions}
\label{zdis}

\subsection{Plane-Parallel Approximation}

In redshift space, the radial coordinate $\s$ of a galaxy is given by
its observed radial velocity, a combination of its Hubble flow plus
``distortions'' due to peculiar velocities. The mapping from
real-space position ${\bf \x}$ to redshift space is given by: 

\beq
\s=\x - f \ \uz(\x) {\hat z},
\label{zmap} 
\eeq 

\noindent where $f(\Omega_m) \approx \Omega_m^{0.6}$ is the logarithmic
growth rate of linear perturbations, and $\u(\x) \equiv -
\v(\x)/({\cal H} f)$, where $\v(\x)$ is the peculiar velocity field,
and ${\cal H}(\tau) \equiv (1/a)(da/d\tau)= Ha$ is the conformal
Hubble parameter (with FRW scale factor $a(\tau)$ and conformal time
$\tau$).  In Eq.~(\ref{zmap}), we have assumed the ``plane-parallel''
approximation, so that the line-of-sight is taken as a fixed
direction, denoted by ${\hat z}$. Using this mapping, the Fourier
transform of the density field contrast in redshift space
reads (\cite{SCF99})

\beq
\ds(\k) = \int  \frac{d^3x}{(2\pi)^3} {\rm e}^{-i \k\cdot\x}
{\rm e}^{i f k_z \uz(\x)} \Big[ \d(\x) + f \nabla_z \uz(\x) \Big].
\label{d_s}
\eeq

\noindent This equation describes the fully non-linear density field
in redshift space in the plane-parallel approximation.  In linear
perturbation theory, the exponential factor becomes unity, and we
recover the well known formula (\cite{Kaiser87})
\beq 
\ds(\k)=\d(\k)\ (1+f\mu^2)
\label{delta_sl}.  
\eeq

Given the general formula, Eq.~(\ref{d_s}), and the local biasing
scheme in Eq.~(\ref{locbias}) (with $\varepsilon(\x)=0$), it is easy
to work out the perturbative solutions order by order
(\cite{SCF99}). Let's write the Fourier components of the density
field in redshift space as

\beqa
\d_s(\k) &=& \sum_{n=1}^\infty D_1^n \int \dD_n\
Z_n(\k_1, \ldots, \k_n)\ \prod_{i=1}^n \d_1(\k_i)\ d^3k_i , \nonumber \\
\eeqa

\noindent where $Z_n$ are dimensionless, symmetric, scalar functions
of their arguments, $D_1$ is the linear growth factor of the density
field, and $\dD_n \equiv \d_D(\k - \k_1\ldots - \k_n)$. In particular
the first and second-order PT kernels are (\cite{HBCJ95,VHMM98,SCF99}) 

\beqa
Z_1(\k) &=& (b_1 + f \mu^2) \\ 
Z_2 (\k_1,\k_2) &=& \frac{b_2}{2} + b_1 F_2 (\k_1,\k_2) 
+f \mu^2 G_2 (\k_1,\k_2) + 
\frac{f \mu k}{2} \Big[ \frac{\mu_1}{k_1} (b_1+f \mu_2 ^2) +
\frac{\mu_2}{k_2} (b_1+f \mu_1 ^2) \Big], 
\label{z2}
\eeqa 

\noindent where we denote $\mu \equiv \k \cdot {\hat z}/k$, with $\k
\equiv \k_1 +\ldots +\k_n$, and $\mu_i \equiv \k_i \cdot {\hat
z}/k_i$. As in Eq.~(\ref{d2}), $G_2$ denotes the second-order kernels
for the velocity-divergence field, given by Eq.~(\ref{F2}) after
replacing $5/7 \rightarrow 3/7$ and $2/7 \rightarrow 4/7$. We can now
write the power spectrum and bispectrum in redshift-space,

\beq
P_s(\k) = Z_1(\k)^2\  P(k),
\eeq
\beq
B_{123} = 2 Z_2(\k_1,\k_2) Z_1(\k_1) Z_1(\k_2)\ P(k_1)
P(k_2) + {\rm cyc.}
\label{zbisp}
\eeq

\noindent Note that both statistics do now depend on the direction of
the wave vectors with respect to the line of sight, since the
redshift-space mapping in the plane-parallel approximation breaks
statistical isotropy. One can proceed to decompose these statistics in
multipole moments; however, in this paper we will be dealing
exclusively with their monopoles. The resulting reduced bispectrum
$Q$, defined as in Eq.~(\ref{q}) in terms of monopole quantities, now
becomes a function of $\beta \equiv f/b_1$, $b_1$ and $b_2$. Even
though Eq.~(\ref{Qg}) is not valid anymore in redshift-space, it is
still a reasonable approximation (\cite{SCF99}). As a result, even
including redshift-distortions, the reduced bispectrum $Q$ is weakly
sensitive to $\Omega_m$ (or $\beta$) (\cite{HBCJ95,VHMM98,SCF99}).

\subsection{Effect of Radial Redshift-Distortions}

A standard procedure with redshift distortions is to treat them in the
plane-parallel approximation (\cite{Kaiser87}). When dealing with
surveys with substantial sky coverage, this approximation is expected
to break down. In an all-sky survey (as \IRAS) the radial character of
redshift distortions mean that clustering statistics are statistically
isotropic but inhomogeneous (rather than statistically homogeneous but
anisotropic, as in the plane-parallel case). The breakdown of
statistical homogeneity means that Fourier modes are not independent
anymore, in particular, the power spectrum is not diagonal,
e.g. different band powers are correlated, even in the linear regime
(\cite{ZaHo96}).

We will show, however, that for the monopole of the power spectrum and
bispectrum in the all-sky case, radial and plane-parallel distortions
agree with each other extremely well. This can be understood from the
fact that the monopole is not sensitive to the orientation of
structures but to the overall spherical average of power. On the other
hand, higher-order multipoles (and thus the full unaveraged over
angles statistics) such as the quadrupole are much more sensitive to
the radial character of the distortions; in fact, in the limit of full
sky, the anisotropy (with respect to a fixed direction) vanishes.

Figure \ref{fig_pkradial} illustrates this for the monopole of the power
spectrum. Solid lines denote the undistorted non-linear power
spectrum, whereas dotted (dashed) lines denote the power spectrum
monopole under radial (plane-parallel) redshift-space mapping. The
radial results overlap with the plane-parallel ones except for a very
small difference at large scales, where the radial results are
slightly smaller. This is indeed expected at the 10\% level for the
scales plotted (\cite{HeTa95}). A similar result holds for the scale
dependence of the bispectrum. In the top left panel of
Fig.~\ref{fig_sys} we show the ratio of the bispectrum under radial
($Q_R$) to plane-parallel ($Q_P$) redshift-space mapping as a function
of the angle $\theta$ between $\k_1$ and $\k_2$ for $k_2=2k_1=0.105$
h/Mpc (triangles) and $k_2=2k_1=0.21$ h/Mpc (squares). In order to get
this result we have averaged over 200 realizations of
$\Lambda$CDM. There is an apparent trend at the few percent level that
indicates that under the radial mapping the configuration dependence
is very slightly suppressed, but the effect is very small even at
these large scales. In the following we use radial distortions,
although for our purposes plane-parallel approximation would have been
a good approximation. We use Fourier modes even though they are not
the natural set of modes anymore in the presence of radial
redshift-distortions (see e.g. \cite{HeTa95}; Hamilton \& Culhane
1996).

\subsection{Dependence of $Q$ on Cosmology and Power Spectrum}

As discussed before, $Q$ depends mainly on the spectral index,
biasing, Gaussianity of initial conditions, and it is quite
insensitive to the matter density $\Omega_m$ and should be independent
to leading order on the power spectrum normalization $\sigma_8$. We
now quantify these dependences in the presence of non-linear redshift
distortions. Figure~\ref{fig_sys} shows the results of averaging 200
realizations of 2LPT, as described before, for different models. The
top right panel shows the dependence of $Q$ on the matter density
parameter $\Omega_m$, for power spectrum shape given by
$\Gamma=0.21$. The scatter plot corresponds to measurements of $Q$ in
{\em all} triangles (4741 in total) between scales of $k=0.05$ h/Mpc
and $k=0.2$ h/Mpc. Equilateral triangles correspond to the small
amplitude region (lower left corner), whereas 
colinear triangles ocupy the upper right corner. As predicted by
tree-level PT, the $\Omega_m=1$ case shows a higher colinear amplitude
and lower equilateral amplitude than the $\Omega_m=0.3$ case, although
the precise value of these amplitudes does not quite agree with
tree-level PT in the presence of non-linear redshift distortions, as
we saw in Fig.~\ref{fig_bktheta}. The effect of a change in $\Omega_m$
is of the order of $10\%$ (Hivon et al. 1995).

The bottom left panel shows the effect of changing $\sigma_8$ for
$\Lambda$CDM models. As we see, non-linear redshift distortions
introduce a very small dependence, with larger $\sigma_8$ leading to a
slightly smaller configuration dependence of $Q$, but the effect is
negligible for the analysis of \IRAS surveys given the error bars
involved, as we shall see. On the other hand, as expected, the
dependence on the shape of the power spectrum, parameterized by
$\Gamma$ in CDM models is quite strong due to a change of the spectral
index with scale. We see from the bottom right panel that for
$\Omega_m=1$, a $\Gamma=0.5$ has a much weaker configuration
dependence than a $\Omega_m=0.21$, as expected since the spectral
index of the latter is more negative at a given scale. The solid line
in this panel shows the relation $Q(\Gamma=0.21)= 1.3
Q(\Gamma=0.5)-0.05$, which fits the results quite well. As a result of
this, a $\Gamma=0.5$ model when compared with a given galaxy
bispectrum will lead to smaller values for the bias parameters $b_1$
and $b_2$ than a $\Gamma=0.21$ model, assuming no other constraint is
imposed.

\section{Mock Catalogs: Selection Functions and Optimal Weighing}

We generate mock catalogs of different surveys by the following
procedure. We use 2LPT to generate density and velocity fields for a
given cosmological model and normalization $\sigma_8$. In most cases
we use $128^3$ particles in a 600 Mpc/h box, for biased galaxy
distributions we use a parent $200^3$ 2LPT realization from which an
$\approx 128^3$ distribution is generated (see below). An observer is
then picked at random with velocity consistent with that of the local
group with respect to the CMB frame. The full (radial) redshift-space
mapping is performed using the velocity field, to obtain the
redshift-space distribution. Given the particle distribution in
redshift-space, ``galaxies'' are selected using the desired IRAS
survey selection function, replicating the box if necessary, including
the galactic cut, and rejecting galaxies closer than 20 Mpc/h from the
observer. The selection functions are given by:

\beq 
n(r) = \frac{A}{r^2} \frac{\left( r/r_0 \right)^{a}}{1+
\left( r/r_0\right)^{b} }, 
\label{selfun}
\eeq

\noindent with parameters given in Table~\ref{sfpar}. These were
kindly provided by H.~Feldman, and obtained by fitting the actual
redshift distribution observed in the different \IRAS surveys. 

In order to correct for the selection function and the geometry of the
survey, we use the standard FKP method (\cite{FKP94}; hereafter
FKP). For each realization of the mock survey, a corresponding random
catalogue is generated using the same geometry and selection function,
but about ten times more dense. We then measure statistics (power
spectrum and bispectrum monopoles) for the auxiliary $F(\k)$ field,

\begin{equation}
F(\k) = \int d^3x\ w_k(\x) \left[n_g(\x)- \alpha n_r(\x) \right] \exp (i
\k \cdot \x),
\label{Ffkp}
\end{equation}

\noindent where $n_g(\x)$ and $n_r(\x)$ are the galaxy and random
catalog density fields, and $\alpha = N_g/N_r$ scales the overall
random density to the survey density. Note that the use of a random
catalog is not necessary, one could just as well use the survey
selection function multiplied by the survey mask to substract the
expected shot noise; or even better, substract the ``actual'' shot
noise using the data itself (Hamilton 1997). The weight function
$w_k(\x) \equiv 1/\left[1+ \bar{n} P(k) \right]$ is chosen as to
minimize the variance of the power spectrum estimator at scales small
compared to the survey size (FKP). In fact, under the same assumptions
plus Gaussian errors, the same weight function minimizes the variance
of the higher-order correlation functions as well. Indeed, the
variance $(\Delta \hat{T}_m)^2$ in the estimator of the $m-$point
spectrum $\hat{T}_m$ in this limit is

\begin{equation}
(\Delta \hat{T}_m)^2 \sim \frac{ \int d^3x\ n^{2m}(\x)
\prod_{i=1}^m \left[ P(k_i)+1/n \right] w_{k_i}^2(\x)}{ \left[ \int
d^3x\ n^m(\x) \prod_{i=1}^m w_{k_i}^2(\x) \right]^2 } ,
\end{equation}

\noindent and taking variations with respect to the weight function
$w_{k_j}(\r)$ it follows that

\begin{equation}
\frac{w_{k_j}(\r) [1+n(\r)P_j] \prod_{i \neq j}^m [1+n(\r)P_i]
w_{k_i}^2(\r)}{\prod_{i\neq j}^m w_{k_i}(\r)} = 
\frac{ \int d^3x \ n^m(\x) \prod_{i=1}^m 
[1+n(\x)P_i] w_{k_i}^2(\x)}{ \int d^3x \ n^m(\x) \prod_{i=1}^m
w_{k_i}(\x) },
\end{equation}

\noindent whose solution is the FKP weight function
$w_k(\x)=1/(1+n(\x) P(k))$ (this result was derived independently by
M.~Zaldarriaga, private communication 2000). In practice however, when
the FKP approximation breaks down, a more complicated procedure must
be used to determine the optimal weight $w_k(\x)$, see e.g. Colombi,
Szapudi \& Szalay (1998) for one-point higher-order statistics. From
Eq.~(\ref{Ffkp}) it follows that

\begin{equation}
P_F(k) = \int d^3k' P_G(|\k-\k'|) P(\k') + (1+\alpha) \int d^3 x w_k(\x)
\bar{n}(\x)^2,
\label{pf}
\end{equation}
 
\noindent where $\bar{n}(\x)=\lexp n_g(\x) \rexp$, $P_F(k)$ and
$P_G(k)$ are the power spectra of the fields $F$ and $G(\x) \equiv
w_k(\x) \bar{n}(\x)$.  For \IRAS surveys, $P_G(k)$ decays quite
strongly with k, as seen in Fig.~\ref{fig_pmock} for QDOT-PSCz
(dashed), 1.2Jy (solid) and 2Jy (dot-dashed), where we assumed for
definiteness $w_k(\x)=1/(1+n(\x) P_0)$ with $P_0=2000$. As a first
approximation we can thus treat $P_{G}(k)$ as a delta function, and we
obtain the power spectrum and bispectrum estimators (Matarrese et
al. 1997) 

\begin{equation}
\hat{P}(k)= \frac{P_F(k)}{I_{22}} - (1+\alpha) \frac{I_{12}}{I_{22}},
\label{Pest},
\end{equation}
\begin{equation}
\hat{B}_{123}= \frac{B_F}{I_{33}} - (\hat{P}_1+\hat{P}_2+\hat{P}_3) 
\frac{I_{23}}{I_{33}}- (1-\alpha^2) \frac{I_{13}}{I_{33}},
\label{Best},
\end{equation}

\noindent where $I_{ab} \equiv \int d^3x\ \bar{n}^a(\x) w_{k_1}(\x)
\ldots w_{k_b}(\x)$ and $\alpha$ denotes the additional shot noise
correction due to the use of a random catalog. In the volume limited
case, $I_{ab}=\bar{n}^a$ and Eqs.~(\ref{Pest}-\ref{Best}) reduce to
the standard shot noise correction results for $\alpha=0$,
$\hat{P}=P-\bar{n}^{-1}$, $\hat{B}=B- \bar{n}^{-1} \sum_i \hat{P}_i -
\bar{n}^{-2}$ (Peebles 1980). The estimator of the reduced bispectrum 
is thus

\begin{equation}
\hat{Q}_{123}= \frac{\hat{B}_{123}}{\hat{P}_1 \hat{P}_2+\hat{P}_2 
\hat{P}_3+\hat{P}_3 \hat{P}_1} \label{Qest}.
\end{equation}

\noindent Note that this involves a non-linear combination of 
estimators, so it is not necessarily and unbiased estimator of $Q$, 
as we shall discuss in the next section.

\section{Effects of Survey Geometry and Sampling}

\subsection{Power Spectrum}

Using Eq.~(\ref{Pest}) for the power spectrum, we find the recovered
power spectrum shown in symbols in Fig.~\ref{fig_pmock}. The mean is
obtained from 785 1.2Jy mock surveys, and the error bars correspond to
the errors scaled to a single realization. The solid line shows the
actual power spectrum in redshift space obtained from 200 2LPT
realizations of $\Lambda$CDM, $\sigma_8=0.7$ (error bars on these
measurements are supressed for clarity, they are completely
negligible). We see from Fig.~\ref{fig_pmock} that the convolution
with the window of the survey $P_{G}(k)$ in Eq.~(\ref{pf}) affects the
recovery of the density power spectrum at scales $k<0.05$ h/Mpc. For
$k>0.05$ h/Mpc, which are the scales we use in this work, the
recovered power spectrum in the narrow-window approximation agrees
very well with the correct answer.

Figure~\ref{fig_pkerror} shows a comparison of the error bars as a
function of scale in the ideal case (2LPT) with those in different
\IRAS surveys. Note that the horizontal scale is now linear, to show
better the range of scales we actually use in this paper, and where
the narrow-window approximation works. The 2LPT realizations are the
same as in the previous figure, having $256^3$ particles in a 1200
Mpc/h box. The Gaussian prediction shown in Fig.~\ref{fig_pkerror} is
obtained by the simple relation $\Delta P(k)/P(k) =2/N(k)$, where
$N(k)$ is the number of modes within a given shell in Fourier space
centered at $k$. From this comparison, we see that even at
$k=0.3$~h/Mpc, the Gaussian approximation to the errors works very
well, consistent with the results from N-body simulations in Meiksin
\& White (1999) and Scoccimarro, Zaldarriaga \& Hui (1999).

Also shown in Fig.~\ref{fig_pkerror} are the corresponding errors for
QDOT, 1.2Jy and PSCz surveys (those of 2Jy are intermediate between
QDOT and 1.2Jy and are supressed for clarity). All the solid lines
correspond to using the FKP weighing $P_0=2000$, constant for all
wavenumbers. The dashed lines show the resulting error bars in PSCz
mock surveys when $P_0=8000$. As expected, the error bars in the
surveys are considerably larger than in the 2LPT case, due to the
smaller volume and the larger shot noise. In fact, when the latter
dominates, errors stop decreasing as $k$ is increased, which happens
first for QDOT, then 1.2Jy and last for PSCz, as expected. The errors
in the PSCz case are smaller for $P_0=8000$ at low $k$ than for
$P_0=2000$ as expected, if one wants to measure longer wavelengths it
is better to weight distant galaxies more (higher $P_0$), as dictated
by the FKP prescription. 

The survey geometry and sparse sampling (shot noise) not only
increases the power spectrum error bars at a given $k$, but also leads
to correlations between different band powers. To quantify this,
Fig.~\ref{fig_pcova} shows the power spectrum correlation coefficient,
obtained from the power spectrum covariance matrix, $C_{ij} \equiv
\lexp (\hat{P}_i-P_i)(\hat{P}_j-P_j) \rexp$, by $r_{ij}\equiv
C_{ij}/\sqrt{C_{ii}C_{jj}}$, for fixed $k_i$ as a function of
$k_j$. The figure shows $k_i=0.053, 0.134, 0.201$ h/Mpc for different
\IRAS surveys and the ``ideal'' 2LPT case. In the latter (bottom left
panel) we see that $r_{ij}$ is practically diagonal, no significant
correlations are induced at these scales by non-linear evolution or
the radial nature of redshift distortions. On the other hand, the 2Jy
shows the slowest decay of $r_{ij}$, as expected due to its small
volume. At the large $k$ end, $r_{ij}$ becomes roughly constant, when
shot noise is expected to dominate the covariance between modes. All
the mock survey results in this figure assume FKP weighing $P_0=2000$,
except for the dashed line in the PSCz case (bottom right panel),
which corresponds to $P_0=8000$. The larger value of $P_0$ in this
case means that we are effectively increasing the volume of the
survey, and thus the amount of cross-correlation between different
band powers decreases. Thus, given the choice between $P_0=2000$ and
$P_0=8000$ at $k \approx 0.1$ h/Mpc where both weights give similar
error bars (see Fig.~\ref{fig_pkerror}), the choice $P_0=8000$ should
be preferred to decrease cross-correlations (which the FKP method does
not attempt to minimize). In general, cross-correlation between
different band powers are not negligible and must be taken into
account when extracting cosmological information from power spectrum
measurements (Eisenstein \& Zaldarriaga 1999; Hamilton 2000; Hamilton
\& Tegmark 2000).

When estimating the power spectrum using likelihood analysis, an
assumption must be made about the likelihood function. At large
scales, the density field can be assumed to be Gaussian, and the power
spectrum appears in the covariance matrix of the Fourier modes. At
scales where non-linear effects become important, $k \approx 0.2-0.3$
h/Mpc, the density field cannot be assumed to be Gaussian anymore. A
complementary approach at these scales is to consider that the power
spectrum distribution about its mean is Gaussian; if there are enough
independent modes contributing to a given shell in Fourier space, by
the central limit theorem the distribution of the power spectrum
should approach Gaussianity. However, due to the finite volume of the
survey and shot noise, different modes are correlated, so it is not
clear a priori at what scales Gaussianity
applies. Figure~\ref{fig_pkpdf} shows the power spectrum probability
distribution function (PDF) obtained from 785 realizations of the
1.2Jy survey. The PDF is plotted as a function of the standarized
variable $\d P/\Delta P$, where $\d P \equiv \hat{P}-P$, $\lexp
\hat{P} \rexp = P$, and $(\Delta P)^2 \equiv \lexp (\hat{P}-P)^2
\rexp$. The four curves show the average power spectrum PDF for
$k<0.067$, $0.067<k<0.134$, $0.134<k<0.201$, and $0.201<k<0.268$ (all
in units of h/Mpc). The solid line denotes a Gaussian distribution,
the actual distribution of power is approximately a $\chi^2$
distribution (with a number of degrees given by the effective number
of modes one would obtain from Fig.~\ref{fig_pkerror} by assuming
$\Delta P/P \approx 2/N_{\rm eff}(k)$) and becomes closer to Gaussian
at small scales, as expected. We see though that at $k=0.268$~h/Mpc
there are still noticeable deviations from Gaussianity.

\subsection{Bispectrum}

\subsubsection{Choice of Triangle Variables}

We now turn to a discussion of the bispectrum in mock surveys,
focusing on the same issues illustrated by the power spectrum results
discussed so far. For each realization of a given survey, the
bispectrum is measured for all triangles with sides between $k=0.05$
h/Mpc and $k=0.2$ h/Mpc. The lower bound is imposed by the scale where
the narrow-window approximation works well, whereas the upper bound is
where shot noise can be best corrected for in a single realization,
still well within the limits of applicability of 2LPT. The triangles
are binned in terms of their sides, $k_1\geq k_2 \geq k_3$, in steps
of $k_f$, i.e. $k_i=10,11,12,...$, where $k_f$ is the ``fundamental''
mode of the Fourier space grid we use, $k_f= 2\pi/1200 \approx 0.005$
h/Mpc. There are 4741 such triangles in total, each of these triangles
in turn contains from $10^4$ to $4\times 10^6$ ``elementary''
triangles (those involving the product of 3 Fourier coefficients).

Since the window function of the survey has a width of order $\Delta
k/k_f \approx 5$ (see Fig.~\ref{fig_pmock}), triangles are very
correlated if their sides differ by less than this width, as a result
the number of ``independent'' triangles is of course much smaller than
4741.  We therefore use coarser bins, defined in terms of shape $s$,
ratio $r$, and scale $k$ parameters (Peebles 1980),

\beq s
\equiv \frac{k_1-k_2}{k_3}, \qquad r \equiv \frac{k_2}{k_3}, \qquad k
\equiv k_3,
\label{Qvar}
\eeq

\noindent where the ``shape'' parameter $s$ obeys $0\leq s \leq 1$
($k_1\geq k_2 \geq k_3$), the ratio $ 1 \leq r \leq 4$, and the
overall scale of the triangle satisfies $10 \leq k/k_f \leq 40$.  The
shape parameter is $s=0$ for isosceles triangles (and if $r=1$ these
are equilateral triangles); for $s=1$ we have colinear triangles.
Using 10 bins in each variable plus the closed triangle constraint
yields 203 triangles. 

One advantage of using these variables is that the main dependence of
$Q$ is on the shape parameter $s$, with a small dependence on the
ratio parameter $r$, and for a scale-free initial power spectrum no
dependence on $k$. Therefore, if we plot $Q$ as a function of a single
variable (``triangle'') which depends on ordered triplets ${s,k,r}$ as
shown in the bottom panel of Fig.~\ref{fig_srk}, $Q$ is roughly an
increasing function of triangle number, which runs from 1 to 203. The
order within the triplet is chosen so $s$ has the slowest variation,
which means that low triangle number corresponds to nearly equilateral
triangles, whereas high triangle number corresponds to nearly colinear
triangles. For example, triangle number $1-10$ corresponds to
equilateral triangles of scales $k/k_f=10,\ldots,40$. The top panel in
Fig.~\ref{fig_srk} shows the reduced bispectrum $Q$ as a function of
triangle number for 2LPT realizations in redshift-space of
$\Lambda$CDM, with $\sigma_8=0.7$. This scheme of using a single
(somewhat complicated) variable that parametrizes $Q$ will be useful
in the following since it allows to deal with all the data at once
(i.e. plot {\em all} the triangles of different shapes and scales in
terms of a single variable).

\subsubsection{Estimator Bias}

The effects of finite volume on higher-order statistics has been well
studied for one-point moments, in both N-body simulations (Colombi,
Bouchet \& Schaeffer 1994, 1995; Colombi, Bouchet \& Hernquist 1996;
Munshi et al. 1999) and galaxy surveys (Szapudi \& Colombi 1996;
Colombi, Szapudi \& Szalay 1998; Kim \& Strauss 1998; Hui \&
Gazta\~naga 1999; Szapudi, Colombi \& Bernardeau 1999). Here we
consider the same problem for three-point statistics, i.e. the
bispectrum.

Figure~\ref{fig_fvol} shows a comparison of the amplitude of the
reduced bispectrum $Q$ as a function of triangle for different surveys
and the underlying bispectrum (labeled 2LPT), to illustrate the
effects of finite survey volume and sparse sampling.  The top panel
shows a comparison of 2LPT with 2Jy mock surveys, we see that the
effects of finite volume (2Jy being the most shallow of the \IRAS
surveys) are rather strong: colinear configurations can be
underestimated by factors as large as $70\%$.  On the other hand, for
the deeper PSCz survey (bottom panel) the finite volume effect is as
expected substantially smaller, at about $20\%$.  The middle panel
compares the QDOT and PSCz surveys, the former being a sparse sample
(one in six galaxies) of the latter.  As we see, apart from a
significant increase in the error bars, sparse sampling does not
significantly affect the amplitude of the reduced bispectrum $Q$
(provided of course that the discretness corrections in
Eqs.~(\ref{Pest}-\ref{Best}) can be performed accurately), except
perhaps by factors of order $10\%$ at colinear configurations.

The finite volume effect is particularly worrisome for the
determination of galaxy bias parameters from surveys, as it has a
similar systematic effect on $Q$ as galaxy biasing with $b_{1}>1$ and
$b_{2}>0$.  For example, if this effect is ignored, one would conclude
erroneously from the unbiased realizations (i.e. $b_1=1$, $b_2=0$) in
Fig.~\ref{fig_fvol} that $b_{1}=2.23, 1.35$ and $b_{2}=2.18, 0.42$ for
2Jy and PSCz respectively.  This ``estimator bias'', the fact that the
estimator of $Q$ is not unbiased, arises because $\hat{Q}_{123}$ in
Eq.~(\ref{Qest}) is a non-linear combination of estimators, which are
nonetheless unbiased (provided that we use the mean value of the
selection function, rather than its actual value to substract the mean
density, and thus avoid the integral constraint bias).  There are two
sources of estimator bias in Eq.~(\ref{Qest}) corresponding to the two
non-linear operations involved; the quadratic combination in the
denominator of Eq.~(\ref{Qest}) and the ratio between numerator and
denominator.  Detailed analysis shows that both contributions turn out
to be important, in particular, the quadratic combination
overestimates the underlying value (obtained from 2LPT) for colinear
configurations (thus leading to an understimate of $Q$ for these
triangles), whereas taking the ratio also leads to an underestimate of
$Q$ for colinear configurations.  In addition, there is a rather small
overestimate of $Q$ for equilateral configurations (only apparent in
the least noisy PSCz mock surveys, bottom panel of
Fig.~\ref{fig_fvol}).  This however can be traced to a small estimator
bias in the bispectrum for equilateral triangles at large scales,
which indicates that the narrow window approximation used to
deconvolve with the survey window bispectrum in Eq.~(\ref{Best}) is
beginning to break down.  However, the effect of this particular
estimation bias is in practice negligible when compared to the other
estimation biases involved.

\subsubsection{Error Bars and Triangle Cross-Correlations}

Figure~\ref{fig_err} shows the relative error on the reduced
bispectrum, $\Delta Q/Q$ as a function of triangle for different
surveys, where $(\Delta Q)^2 \equiv \lexp (Q-\bar{Q})^2 \rexp$. The
top panel shows the ideal 2LPT case, and the errors expected in a
single realization of the 2Jy survey. Similarly, the middle and bottom
panels show the corresponding results for 1.2Jy and PSCz surveys. We
see that in general errors decrease as configurations become more
colinear (simply because the number of triangles increases), with
spikes due to shot noise at small scales. For example, the spike at
the tenth triangle is due to equilateral triangles at small scales
($k=0.2$ h/Mpc) becoming dominated by shot noise. For QDOT (not
shown), $\Delta Q/Q> 1$ for all triangles, with most triangles having
significantly larger errors. In Fig.~\ref{fig_chisqm}, we show similar
results for $\chi^2$ initial conditions. The 2LPT results are from
Scoccimarro (2000), and the 1.2Jy results show the rather strong 
finite volume effects (top panel) and the increase error
bars with respect to the Gaussian initial conditions case (compare
with middle panel in Fig.~\ref{fig_err}). These results are noiser
than in the Gaussian case shown in Figs.~\ref{fig_fvol}-\ref{fig_err}
because only 200 (instead of $\approx 10^3$) mock catalogs are used to
compute $Q$ and $\Delta Q$.

Figure~\ref{fig_Qcova} shows the bispectrum correlation coefficient,
obtained from the covariance matrix as in the power spectrum
case. This measures correlations between different triangles and shows
that in the ideal case (2LPT, top panel) different triangles are
practically independent, whereas when survey geometry and sampling are
included (1.2Jy, bottom panel), correlations between different
triangles are significant. The peaks in the correlation coefficient
can be matched with triangles that have their sides of similar length,
as expected from the power spectrum correlations shown in
Fig.~\ref{fig_pcova}.

\subsubsection{The Distribution of $Q$}

The main ingredient to implement a likelihood analysis is to
understand the probability distribution (PDF) of the reduced
bispectrum $Q$. In order to do that, we use the 2LPT Monte Carlo
realizations to construct the PDF's for different mock surveys. As
found for the power spectrum (Fig.~\ref{fig_pkpdf}), we expect that
bispectrum correlations due to survey geometry and sampling as shown
in Fig.~\ref{fig_Qcova} invalidate straightforward application of the
central limit theorem, and thus convergence to a Gaussian distribution
for $Q$.

Figure~\ref{fig_Qpdf} shows the results. We found that the
distributions of $Q$ for all the triangles we consider were very
similar when plotted in terms of the scaled variable $\d Q/\Delta Q$,
where $\d Q \equiv Q- \bar{Q}$. We have thus averaged all these PDF's
to increase the signal to noise. In reality one expects these PDF's to
approach Gaussianity when the error is small (i.e. survey geometry and
sampling are not significant), and thus a variation of PDF on scale;
however for the range of scales we consider the errors don't change
significantly (see Fig.~\ref{fig_err}), so to first approximation the
PDF's do not change. 

The dotted lines in the top panel of Fig.~\ref{fig_Qpdf} shows indeed
that in the ideal case (2LPT) the PDF of $Q$ approaches Gaussianity
(solid smooth line). On the other hand, mock surveys show clear
deviations from Gaussianity, with positive skewness, exponential tails
for 1.2Jy and 2Jy surveys, and in the sparse sampling case (QDOT)
power-law tails and rather strong kurtosis. For $\chi^2$ non-Gaussian
initial conditions (bottom panel), the non-Gaussianity is even more
pronounced, as expected. This clearly shows that likelihood analysis
for current surveys based on a Gaussian likelihood for the bispectrum
are not justified. In particular, the skewness of the PDF can lead to
a ``statistical'' bias in the estimation of galaxy bias: since the
most likely value for $Q$ is to underestimate the mean, use of a
Gaussian likelihood will infer bias parameters that  overestimate 
$b_1$ and understimate $b_2$.

\section{Likelihood Analysis}

\subsection{$Q$ Eigenmodes}

A general treatment of likelihood non-Gaussianity and
cross-correlations between triangles is very complicated. Here we will
make the assumption that the eigenmodes of $Q$, defined by
diagonalizing its covariance matrix, are effectively independent
(which would be true if their joint PDF were Gaussian), and thus write
the full likelihood as a product of eigenmode-likelihoods (which are
computed by 2LPT realizations of a given survey, and are generally
non-Gaussian). In general, diagonalizing the covariance matrix does
not guarantee that the eigenmodes are independent (third and
higher-order correlations could be still non-zero), but as we shall
see this simplification seems to work very well in practice.

Given a set of measured reduced bispectrum amplitudes 
$ \{Q_{m}\} $, $m=1,\ldots,N_{T}$, where $N_{T}$ is the number 
of closed triangles in the survey, we diagonalize their covariance 
matrix so the $Q$-eigenmodes $\hat q_{n}$,

\beq
\hat{q}_n = \sum_{m=1}^{N_T} \gamma_{mn}\ \frac{Q_m-\bar{Q}_m}{\Delta 
Q_m},  
\eeq
satisfy
\beq 
\lexp \hat{q}_n \hat{q}_m \rexp = \lambda_n^2 \, \d_{nm}, 
\eeq

\noindent where $\bar{Q} \equiv \lexp Q \rexp$ and $ (\Delta
Q)^2\equiv \lexp Q^2 \rexp -\bar{Q}^2$.  These $Q$-eigenmodes have
``signal to noise'' ratio $S/N$,

\beq 
\left( \frac{S}{N} \right)_n \equiv \frac{1}{\lambda_n} \, \left| 
\sum_{m=1}^{N_T} \gamma_{mn} \, \frac{\bar{Q}_m}{\Delta Q_m} \right|.
\label{ston}
\eeq

The physical interpretation of $Q$-eigenmodes becomes clear when
ordered in terms of their signal to noise, as shown in the top panel
in Fig.~\ref{fig_eigen}.  The best eigenmode (highest signal to noise,
$S/N \approx 3$), say $n=1$, corresponds to all weights
$\gamma_{m1}>0$; that is, it represents the overall amplitude of the
bispectrum (Fig.~\ref{fig_eigen}, middle panel).  The next
$Q$-eigenmode, $n=2$ with $S/N \approx 1$, has $\gamma_{m2}>0 $ for
nearly colinear triangles and $\gamma_{m2}<0$ for nearly equilateral
triangles (Fig.~\ref{fig_eigen}, bottom panel); that is, it represents
the configuration dependence of the bispectrum.  Higher-order
eigenmodes contain further information such as variations of amplitude
and shape with scale, but generally have $S/N < 1$ in \IRAS catalogs
(although there are many of them). The $S/N$ of eigenmodes for a
particular survey is sensitive mostly to the biasing scheme and the
amplitude of fluctuations $\sigma_8$, stronger clustering leading to
larger $S/N$ overall.

\subsection{Recovering Bias Parameters}

As discussed above, we can write down the likelihood as a function of
the bias parameters,

\beq
{\cal L}(\alpha_{1},\alpha_{2}) \propto \prod_{i=1}^{N_T}
P_{i}[\nu_i(\alpha_{1},\alpha_{2})],
\label{like}
\eeq

\noindent where $\alpha_{1}\equiv 1/b_1$, $\alpha_{2} \equiv
b_2/b_1^2$, and 

\beq \nu_i(\alpha_{1},\alpha_{2}) \equiv
\frac{1}{\lambda_{i}}\ \sum_{j=1}^{N_T} \gamma_{ji} \, \frac{Q_j -
(\alpha_{1} \bar{Q}_j+\alpha_{2})}{\Delta Q_j},
\label{nu}
\eeq

\noindent where the $Q_j$ ($j=1,\ldots,N_{T}$) are the data, and the
mean $\bar{Q}_j$, standard deviation $\Delta Q_j$, and the
non-Gaussian PDF's $P_{i}(\nu_{i})$ are extracted from the mock
catalogs. Essentially, the $Q$-eigenmode PDF's look similar to those
shown in Fig.~\ref{fig_Qpdf} for $n=1$ (for which $\gamma_{m1}>0$),
and a symmetrized version of it for other eigenmodes where
$\gamma_{mn}$ takes positive and negative values equally frequent. We
assume that to first approximation, $\Delta Q_j$ does not depend on
biasing, which would be true if biasing is linear and
deterministic. Thus, in the following all results shown ignore a
possible dependence of $\Delta Q_j$ on non-linear biasing.

\begin{figure}[t!]
\centering
\centerline{
\epsfxsize=18truecm\epsfysize=18truecm\epsfbox{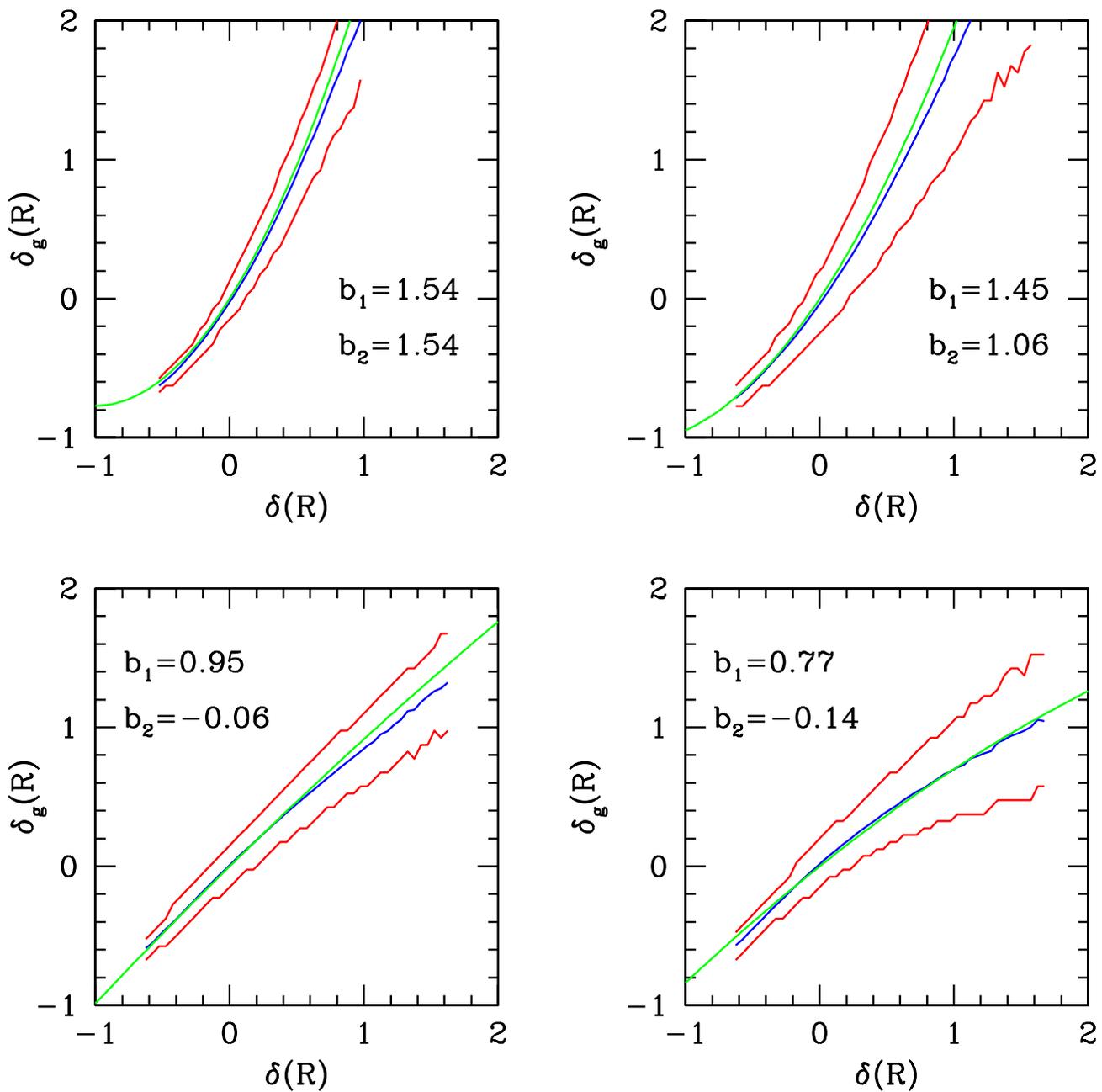}}  
\caption{Recovering stochastic non-linear bias from bispectrum
measurements. The smooth solid line is the bias relation obtained from
the bispectrum with parameters $b_1$ and $b_2$ as shown in each
panel. The true bias relation is represented by the central line
(mean) and the $90\%$ scatter around it. The smoothing scale is $R=10$
Mpc/h .}
\label{fig_bias}
\end{figure}

\begin{figure}[t!]
\centering
\centerline{\epsfxsize=18truecm\epsfysize=18truecm\epsfbox{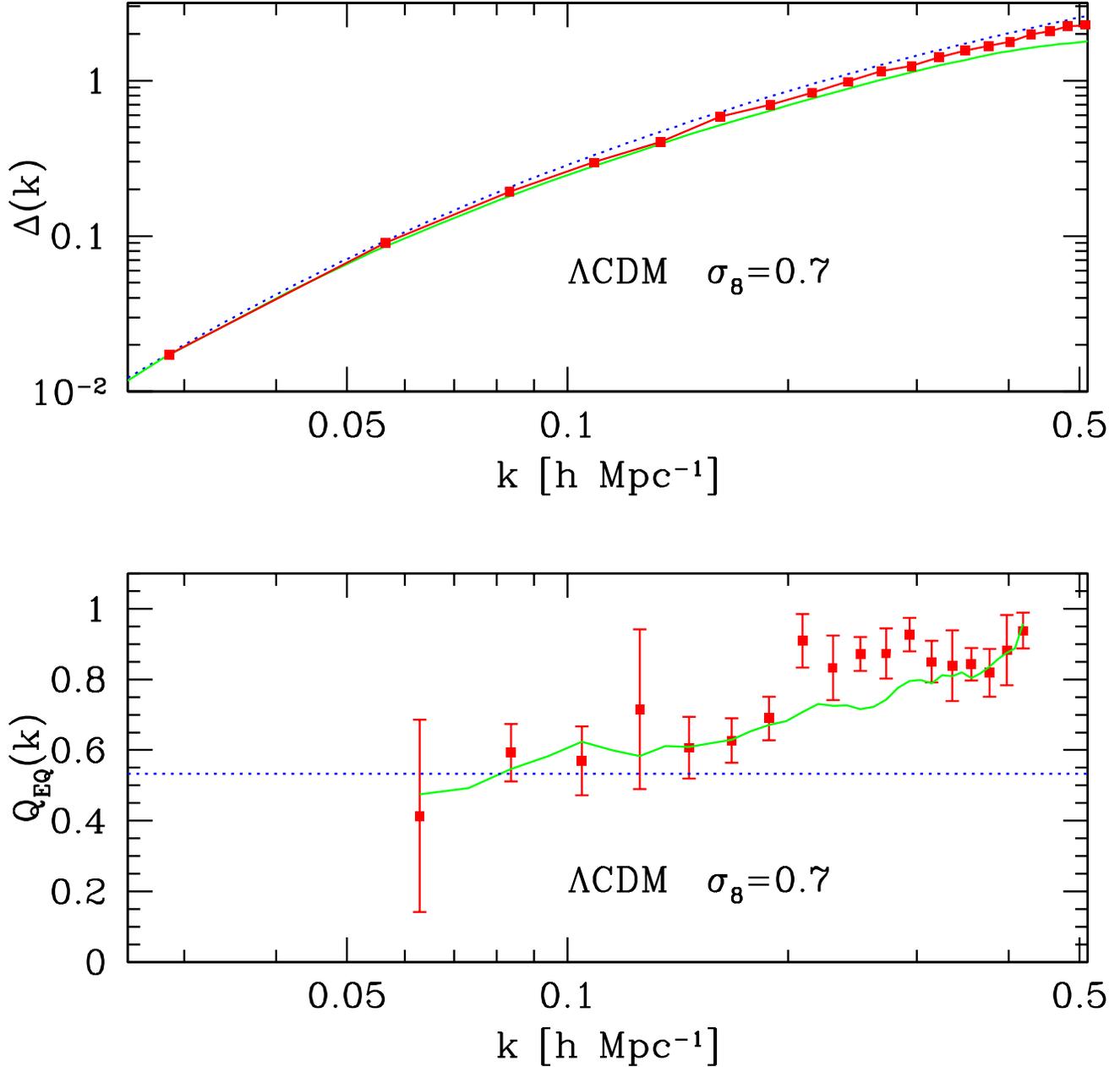}}
\caption{The top panel shows the power spectrum in redshift space as a
function of scale for linear PT (dotted), 2LPT (solid), and N-body
simulations (symbols). The bottom panel shows the reduced bispectrum
for equilateral triangles in redshift space as a function of scale in
tree-level PT (dotted), 2LPT (solid) and N-body simulations
(symbols, with error bars from 4 different realizations).}
\label{fig_sc}
\end{figure}

\begin{figure}[t!]
\centering
\centerline{\epsfxsize=18truecm\epsfysize=18truecm\epsfbox{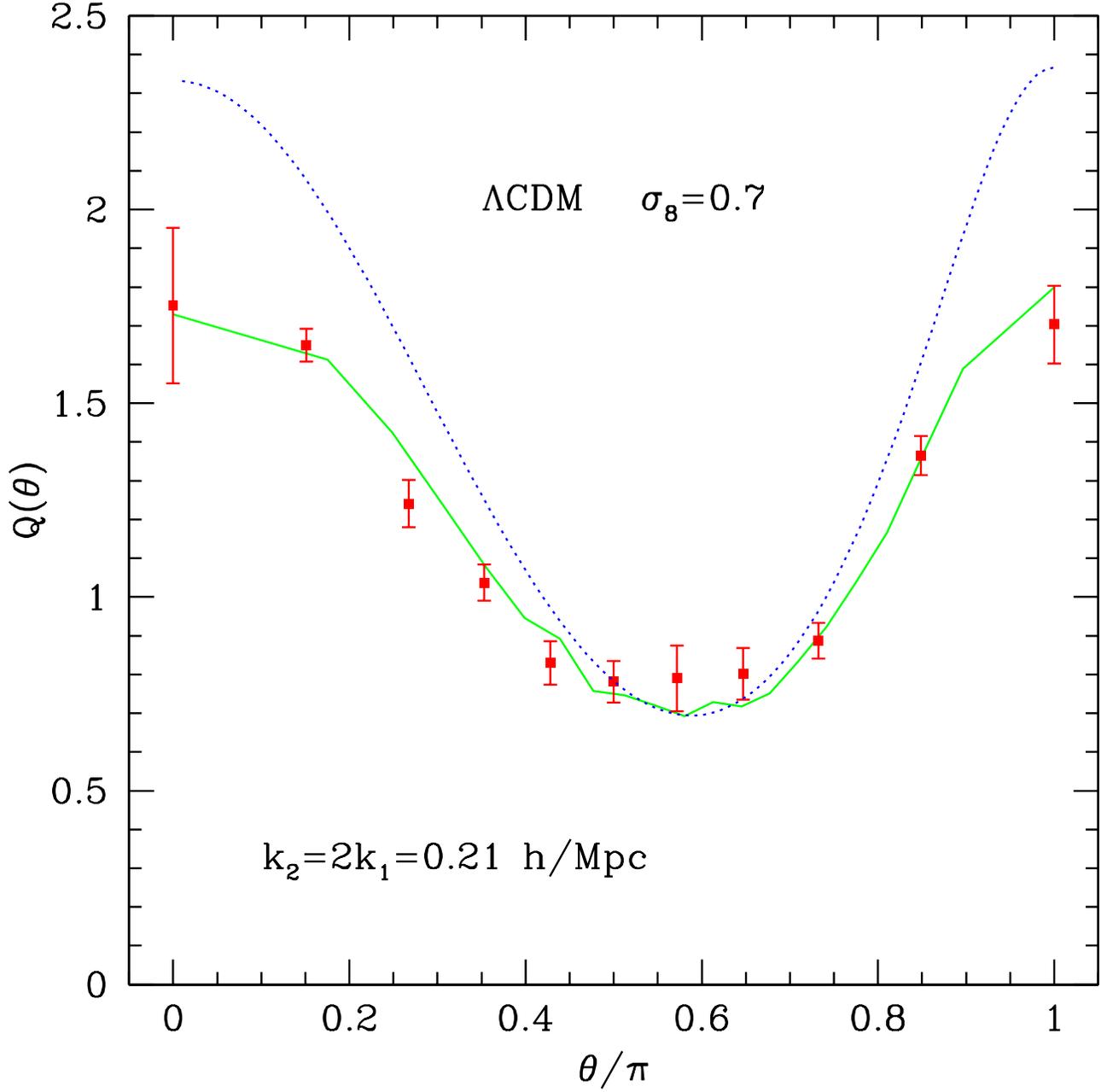}}
\caption{The bispectrum in redshift space for configurations with
$k_2=2k_1=0.21$ h/Mpc as a function of the angle $\theta$ between
$\k_1$ and $\k_2$. Dotted lines denote tree-level PT, solid lines
correspond to 2LPT, and symbols with error bars show the result of 4
realizations of N-body simulations.}
\label{fig_bktheta}
\end{figure}

\begin{figure}[b!]
\centering
\centerline{\epsfxsize=18truecm\epsfysize=18truecm\epsfbox{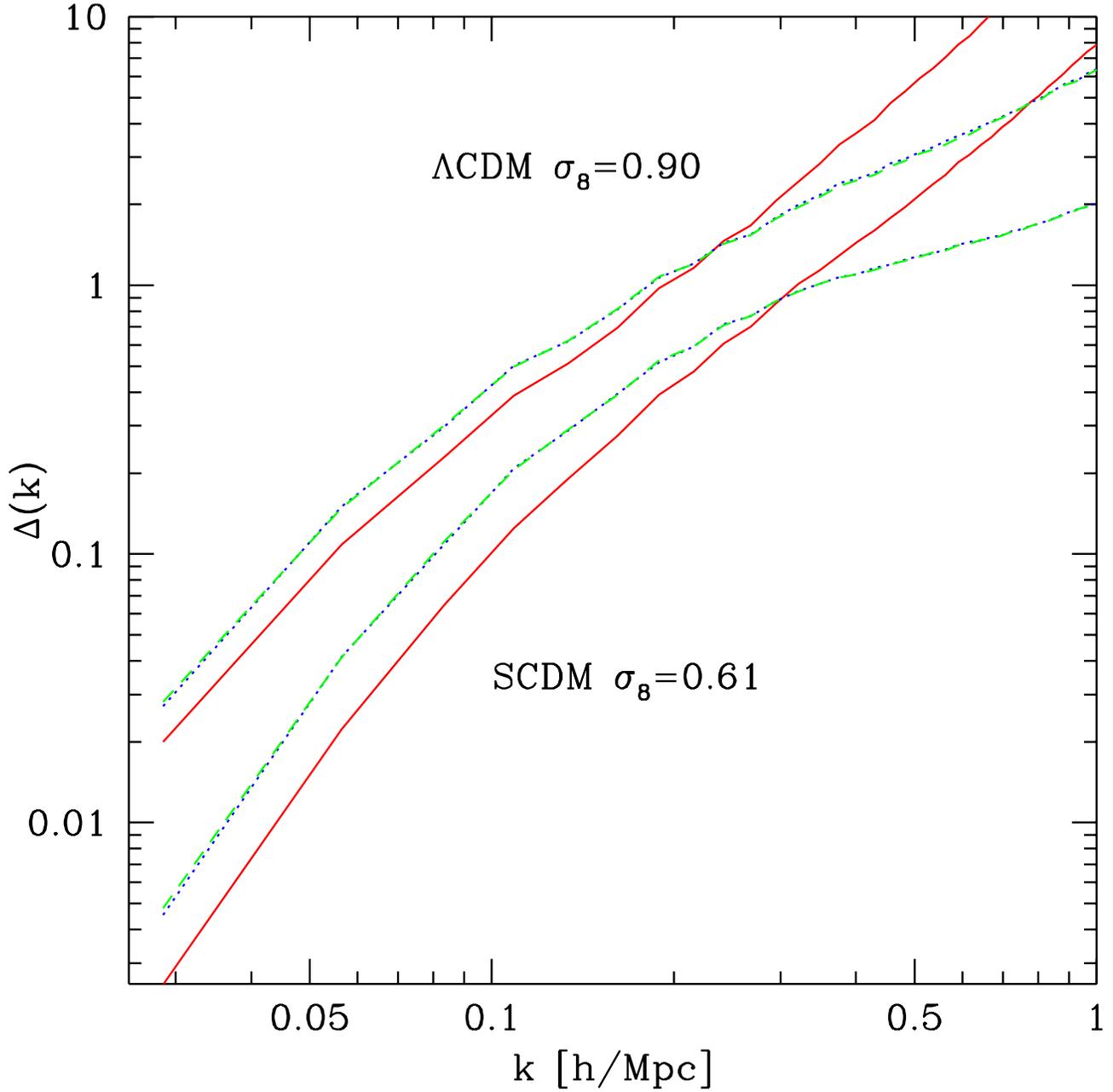}}
\caption{The power spectrum in N-body simulations for $\Lambda$CDM
(top set) and SCDM (bottom). Solid lines denote measurements in real
space, whereas dotted lines and dashed lines (practically on top of
each other) denote measurements in redshift space under radial mapping
and in the plane-parallel approximation, respectively.}
\label{fig_pkradial}
\end{figure}

\begin{figure}[t!]
\centering
\centerline{\epsfxsize=18truecm\epsfysize=18truecm\epsfbox{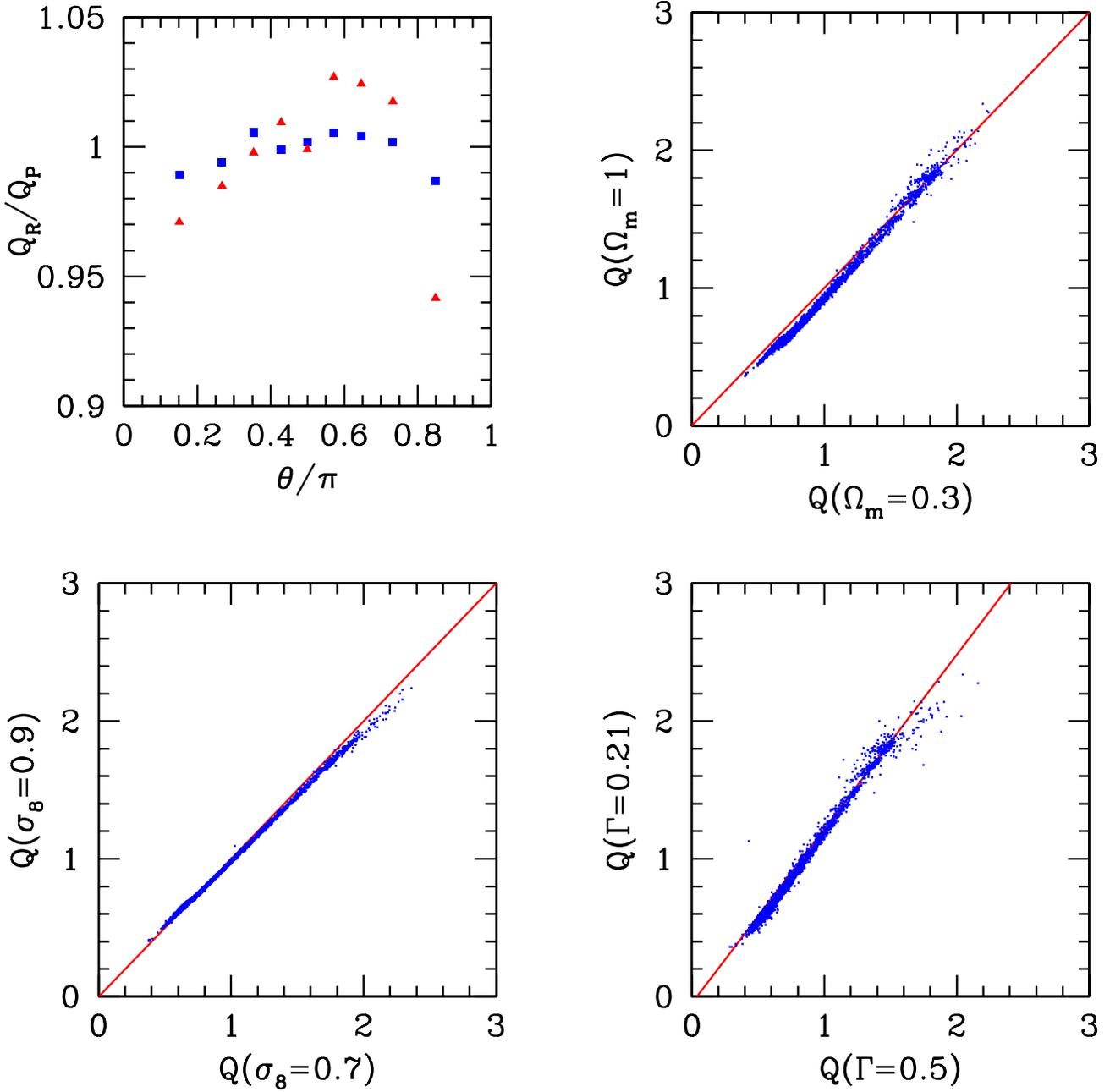}}
\caption{The top left panel shows the ratio of bispectra for radial to
plane-parallel redshift-space mapping as a function of angle as in
Fig.~\protect\ref{fig_pkradial}. Triangles denote $k_2=2k_1=0.105$
h/Mpc, squares $k_2=2k_1=0.21$ h/Mpc. The remaining panels show
sensitivity of $Q$ to change in parameters: $\Omega_m$ (top right),
$\sigma_8$ (bottom left), $\Gamma$ (bottom right).}
\label{fig_sys}
\end{figure}

\begin{figure}[t!]
\centering
\centerline{\epsfxsize=18truecm\epsfysize=18truecm\epsfbox{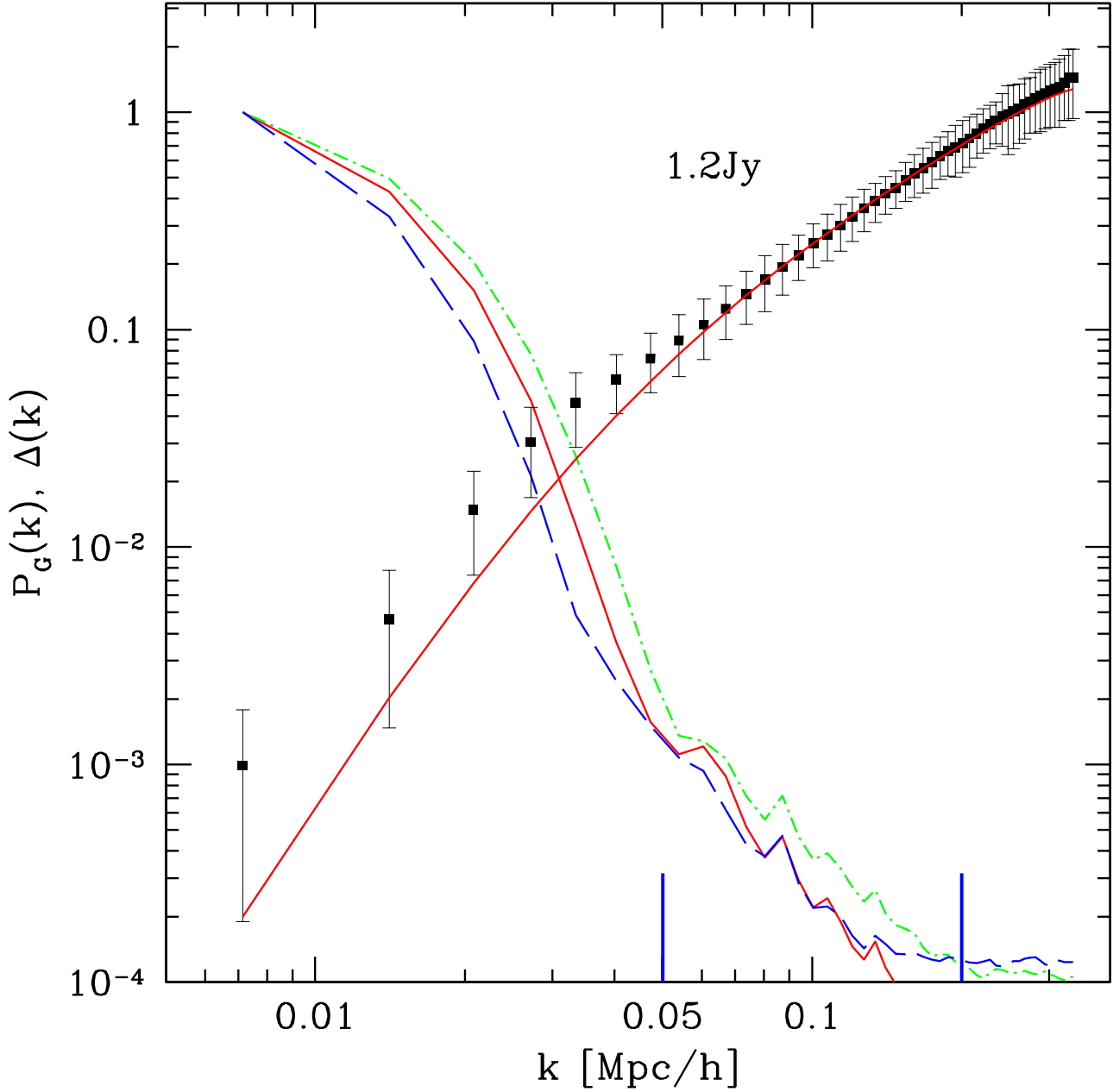}}
\caption{Symbols with error bars (scaled to a single survey) denote
the recovered power spectrum from $\Lambda$CDM realizations of the
1.2Jy survey. Band powers are correlated as shown in top right panel
in Fig.~\protect\ref{fig_pcova}.  The overestimate at scales $k<0.05$
h/Mpc is due to ignoring the width of the survey window.  The plot
also shows the power spectrum of the survey window for 2Jy
(dot-dashed), 1.2Jy (solid) and QDOT-PSCz (dashed).}
\label{fig_pmock}
\end{figure}

\begin{figure}[t!]
\centering
\centerline{\epsfxsize=18truecm\epsfysize=18truecm\epsfbox{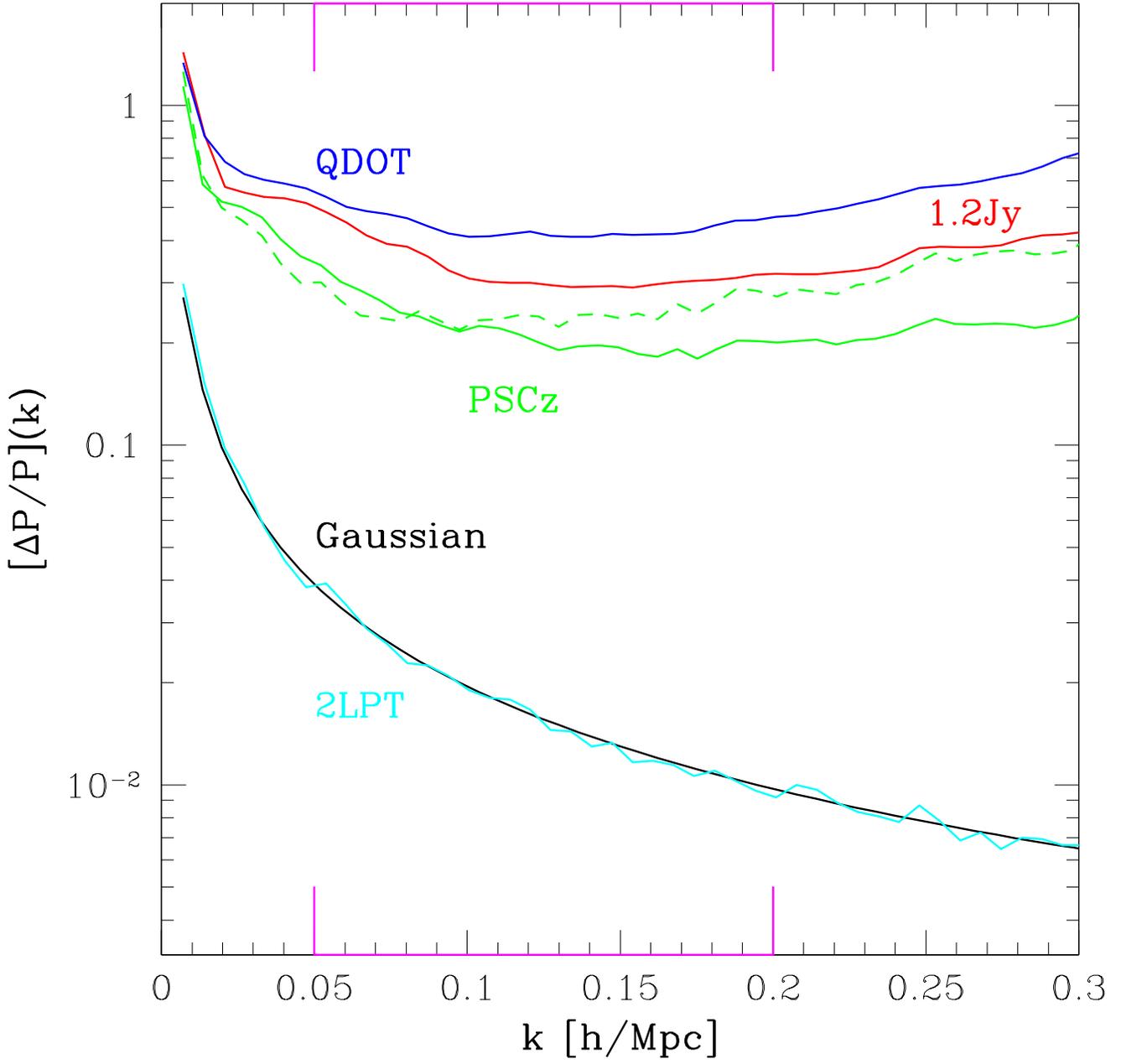}}
\caption{Power spectrum error bars as a function of scale, for 2LPT,
PSCz with $P_0=2000$ (solid) and $P_0=8000$ (dashed), 1.2Jy and 2Jy.}
\label{fig_pkerror}
\end{figure}

\begin{figure}[t!]
\centering
\centerline{\epsfxsize=18truecm\epsfysize=18truecm\epsfbox{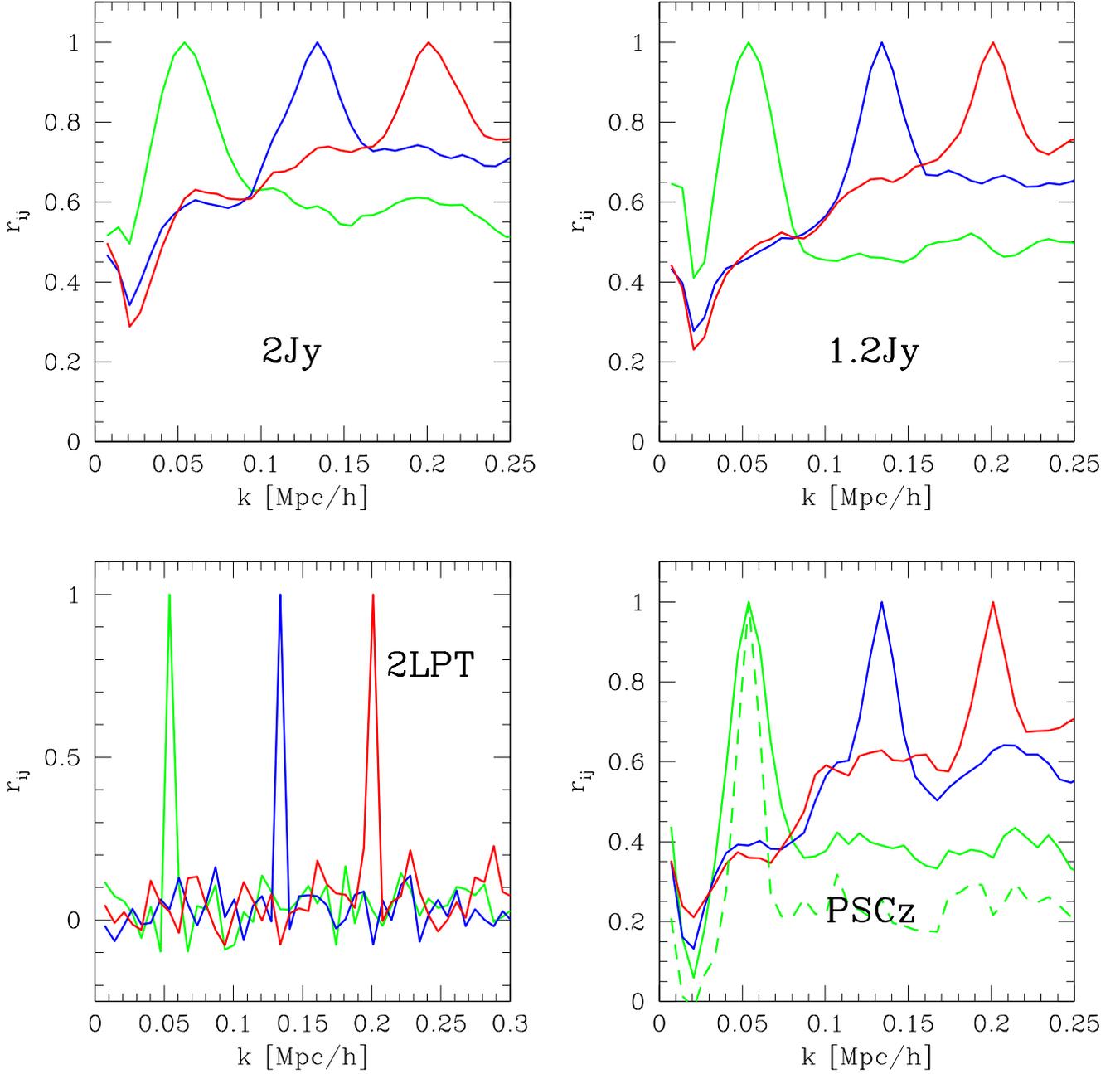}}
\caption{Power spectrum correlation coefficient, for 2Jy, 1.2Jy, PSCz
($P_0=2000$, solid), PSCz ($P_0=8000$, dashed) and 2LPT, as a function
of scale.}
\label{fig_pcova}
\end{figure}

\begin{figure}[t!]
\centering
\centerline{\epsfxsize=18truecm\epsfysize=18truecm\epsfbox{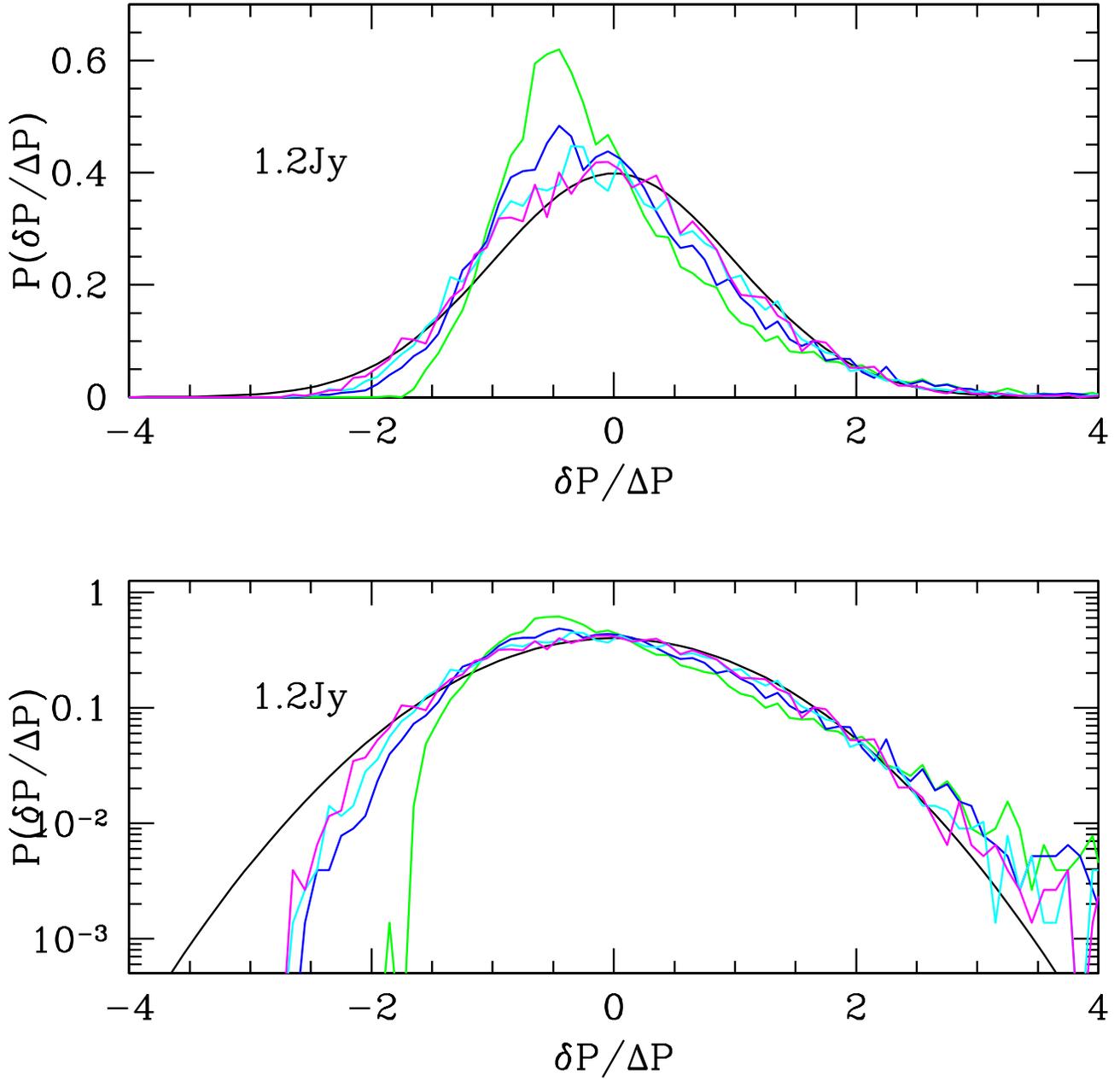}}
\caption{Power spectrum PDF as a function of scale in linear (top) and
logarithmic (bottom) scale, smooth solid line denotes a Gaussian
distribution. From least to most non-Gaussian, scales are
$k/k_f=1-10$, $k/k_f=11-20$, $k/k_f=21-30$, $k/k_f=31-40$, where
$k_f=0.005$ h/Mpc.}
\label{fig_pkpdf}
\end{figure}

\begin{figure}[t!]
\centering
\centerline{\epsfxsize=18truecm\epsfysize=18truecm\epsfbox{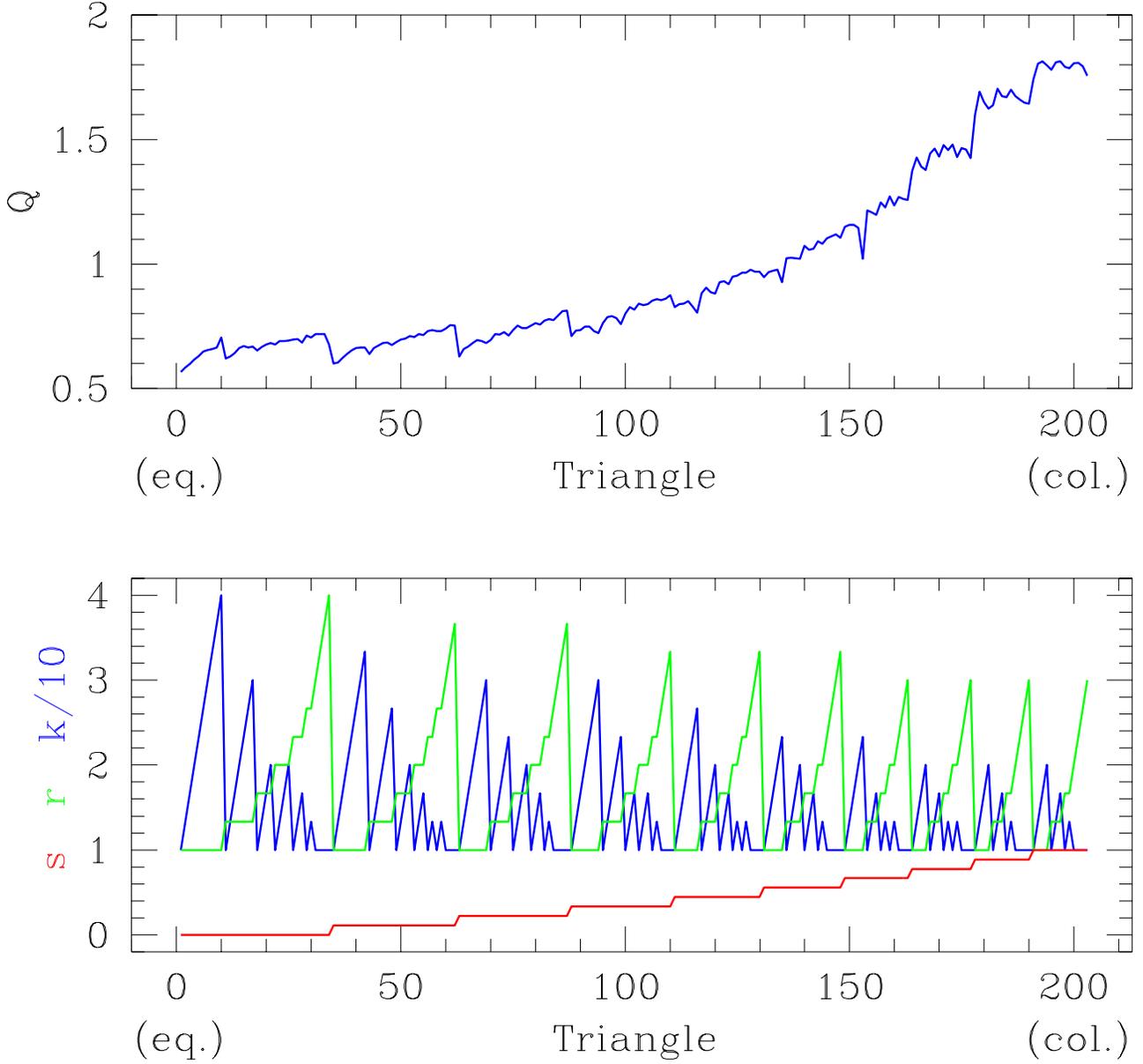}}
\caption{the top panel shows $Q$ measured in the 2LPT realizations for
all triangles, as a function of triangle shape in redshift space, for
$\Lambda$CDM with real space $\sigma_8=0.70$. Bottom panel shows the
parameters $s,r,k$ as a function of triangle number. Essentially,
triangles start equilateral and become colinear as triangle number
increases.}
\label{fig_srk}
\end{figure}

\begin{figure}[t!]
\centering
\centerline{\epsfxsize=18truecm\epsfysize=18truecm\epsfbox{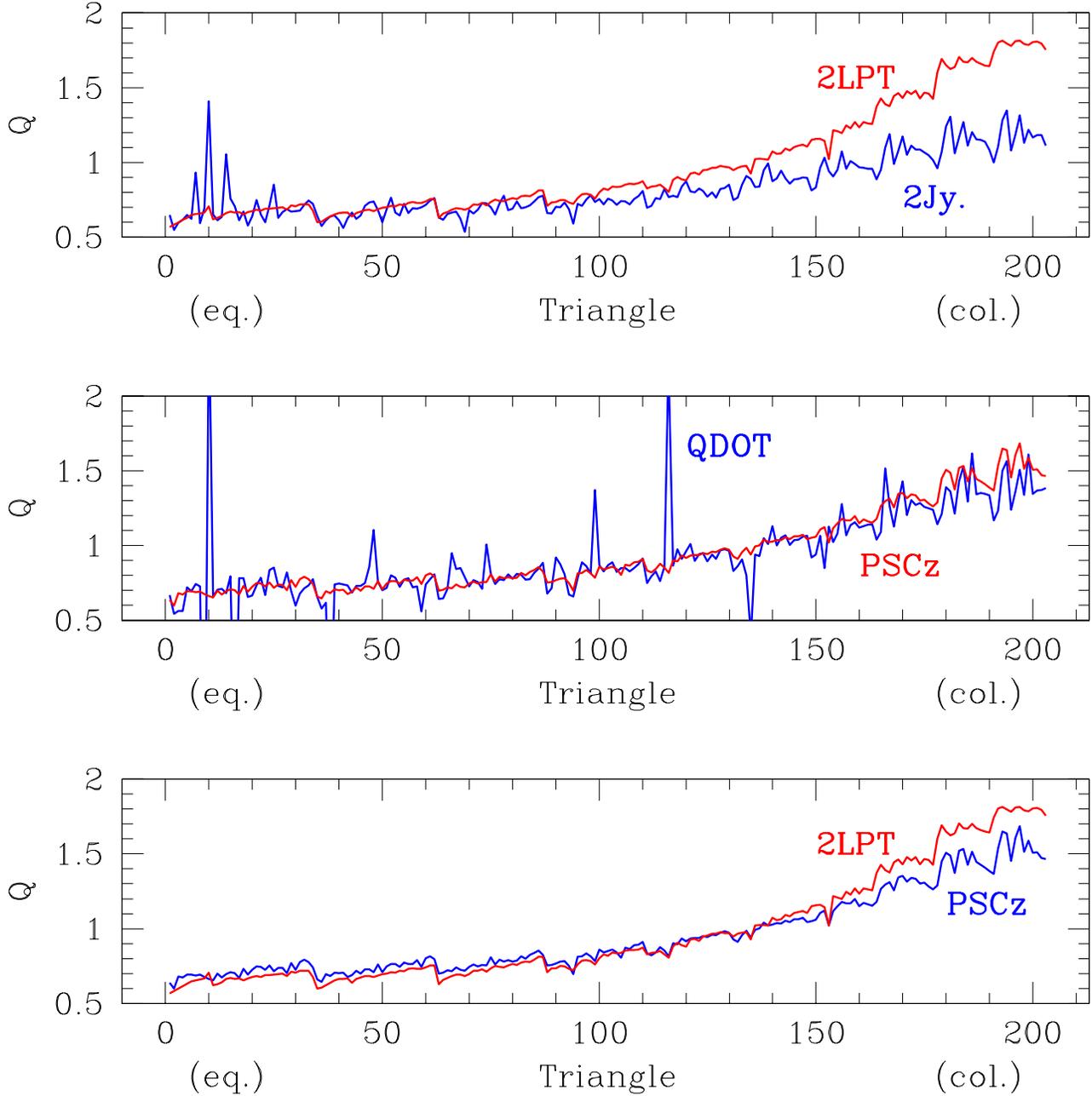}}
\caption{Comparison between underlying bispectrum (2LPT) and the
bispectrum obtained in mock surveys, for 2Jy (top) and PSCz (bottom).
The finite volume of the survey causes estimation bias. The middle
panel shows a comparison between PSCz and QDOT to quantify the effects
of sparse sampling.  }
\label{fig_fvol}
\end{figure}

\begin{figure}[t!]
\centering
\centerline{\epsfxsize=18truecm\epsfysize=18truecm\epsfbox{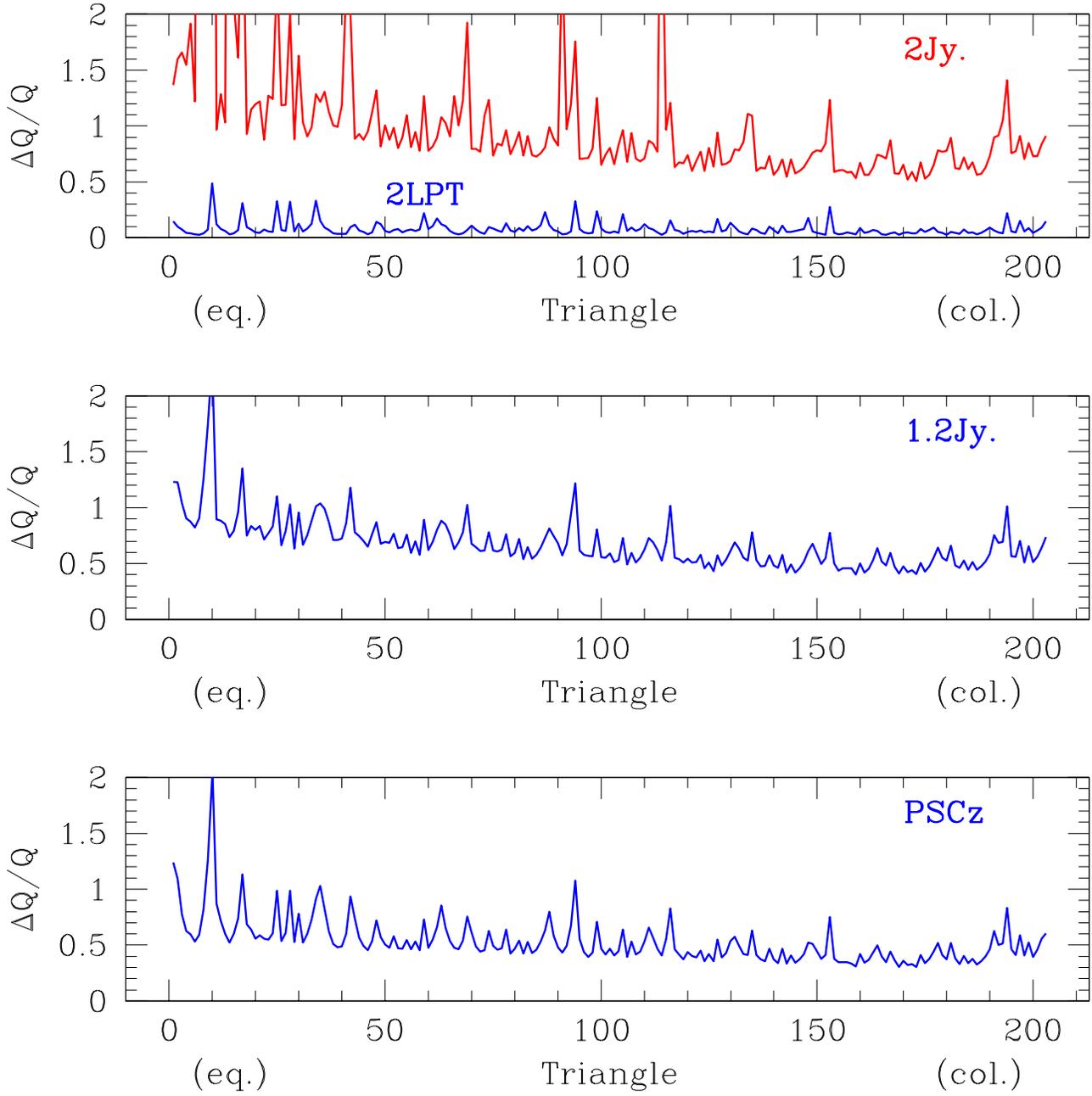}}
\caption{Bispectrum error bars as a function of triangle for different
surveys and 2LPT.}
\label{fig_err}
\end{figure}

\begin{figure}[t!]
\centering
\centerline{\epsfxsize=18truecm\epsfysize=18truecm\epsfbox{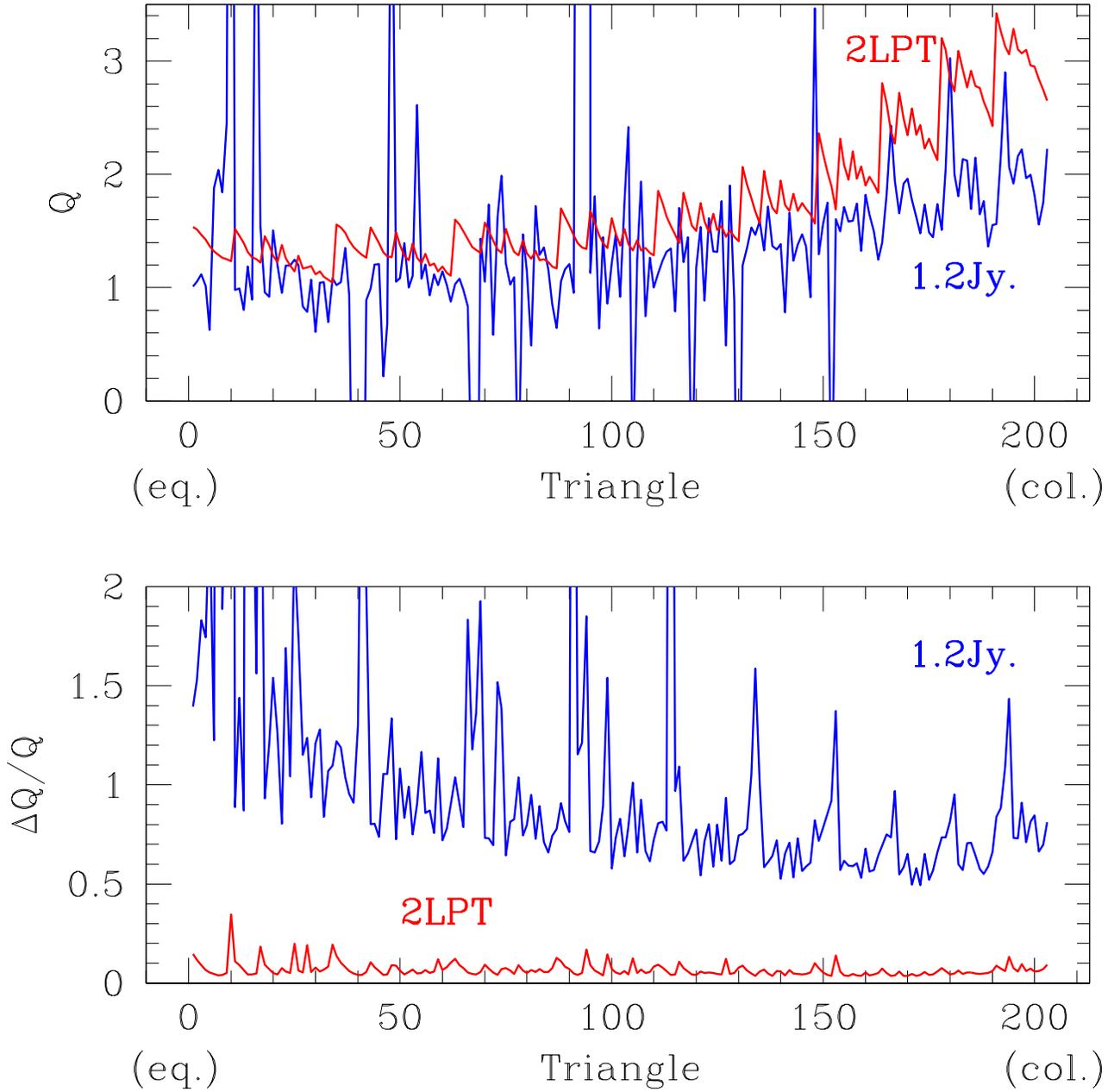}}
\caption{Same as previous two figures for $\chi^2$ initial conditions.}
\label{fig_chisqm}
\end{figure}

\begin{figure}[t!]
\centering
\centerline{\epsfxsize=18truecm\epsfysize=18truecm\epsfbox{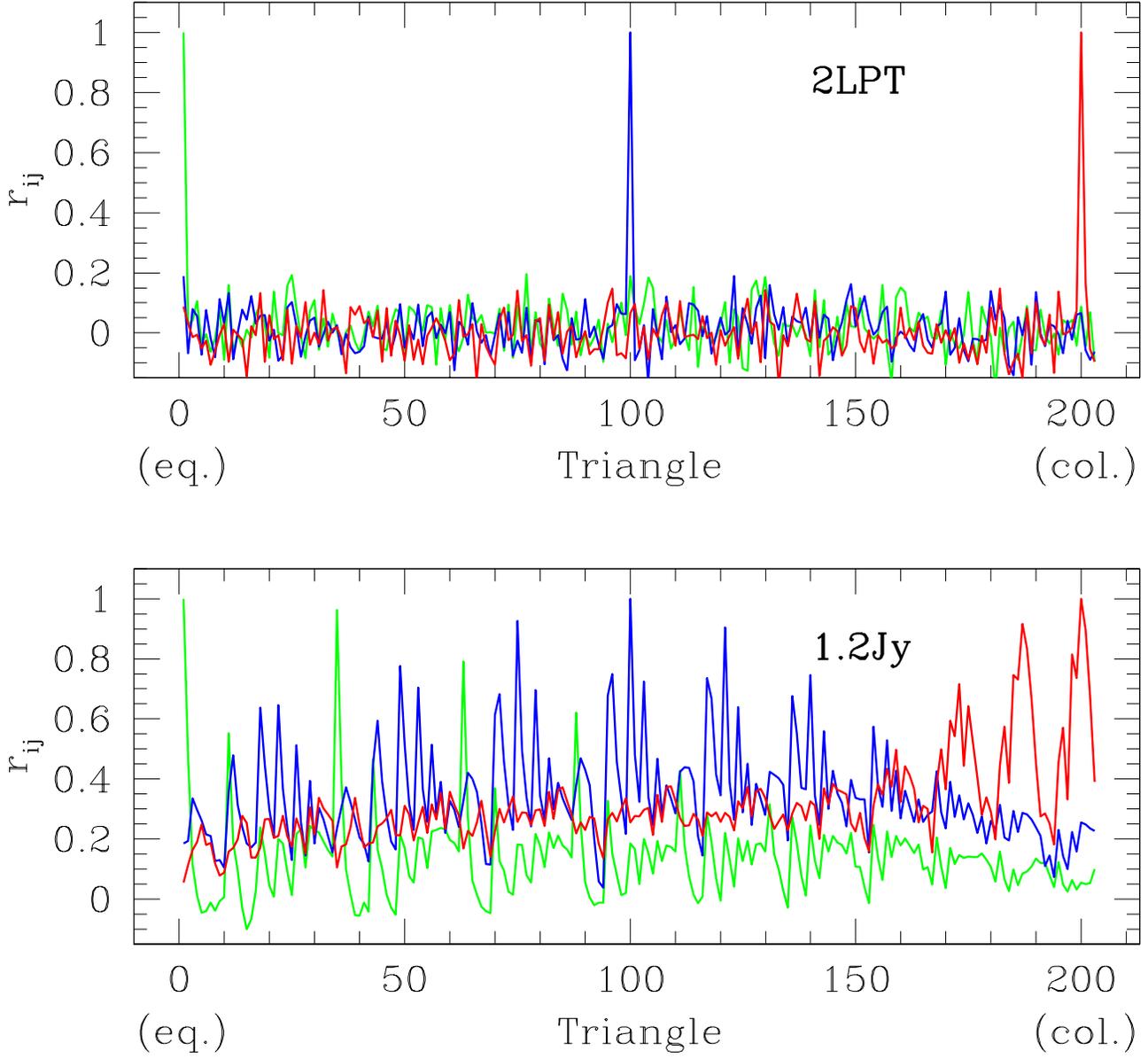}}
\caption{Bispectrum correlation coefficient, as a function of triangle
for 2LPT (top) and 1.2Jy mock surveys (bottom).}
\label{fig_Qcova}
\end{figure}

\begin{figure}[t!]
\centering
\centerline{\epsfxsize=18truecm\epsfysize=18truecm\epsfbox{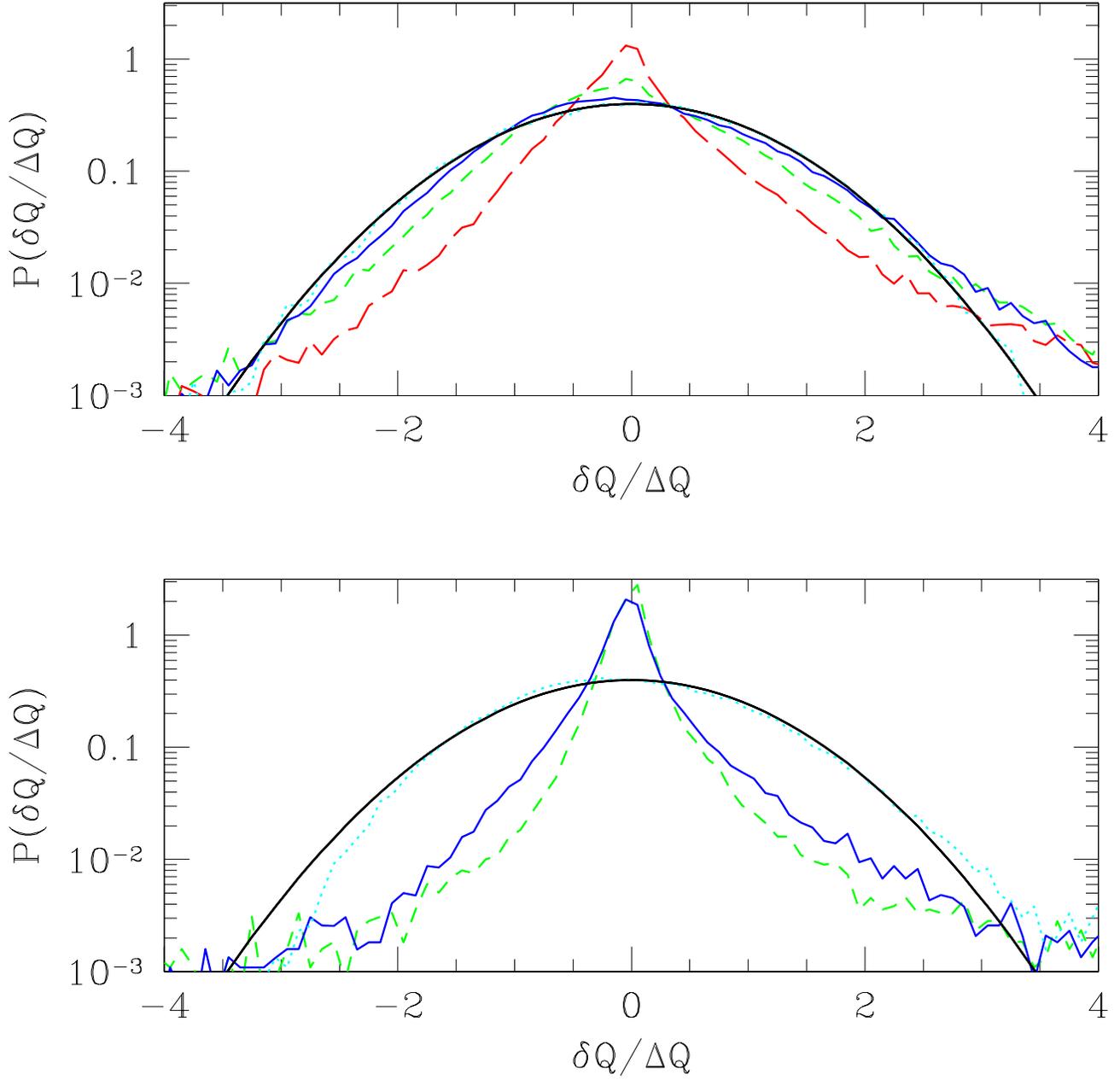}}
\caption{PDF of $\d Q/\Delta Q \equiv (Q-\bar{Q})/\Delta Q $ for
different surveys in models with Gaussian (top) and $\chi^2$
non-Gaussian (bottom) initial conditions: 2LPT (dotted), \IRAS 1.2Jy
(solid), \IRAS 2Jy (dashed), \IRAS QDOT (long-dashed). The smooth
solid curve is a Gaussian distribution.}
\label{fig_Qpdf}
\end{figure}

\begin{figure}[t!]
\centering
\centerline{\epsfxsize=18truecm\epsfysize=18truecm\epsfbox{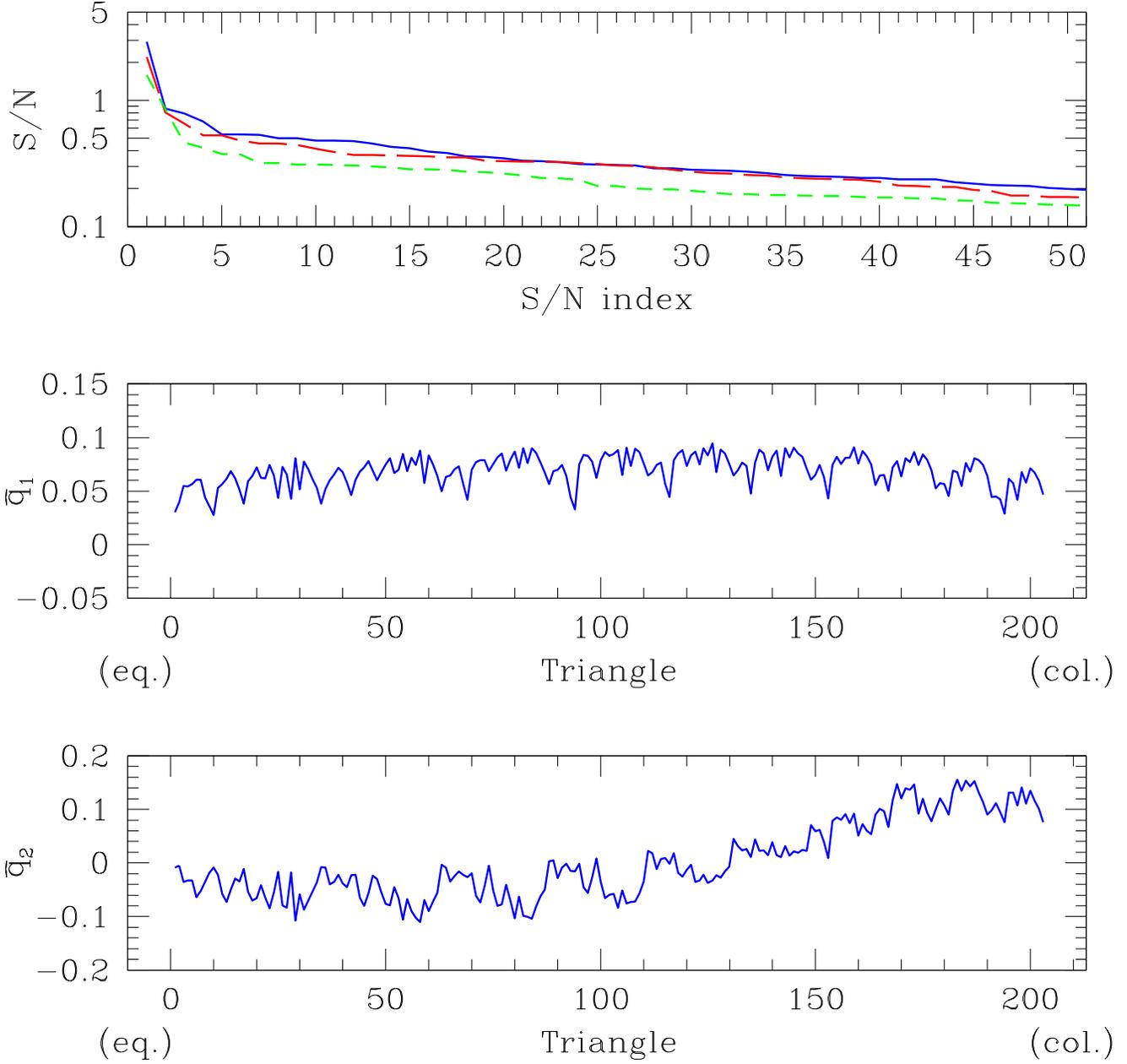}}
\caption{The top panel shows the signal to noise of $Q$-eigenmodes
[see Eq.~(\ref{ston})], with line styles as in
Fig.~\protect\ref{fig_Qpdf}. The remaining panels show the
eigenfunctions for the best determined eigenmode ($\hat{q}_1$),
sensitive to the overall amplitude of $Q$, and the the second best
eigenmode ($\hat{q}_{2}$), which is most sensitive to the
configuration dependence of $Q$.}
\label{fig_eigen}
\end{figure}

\begin{figure}[t!]
\centering
\centerline{\epsfxsize=18truecm\epsfysize=18truecm\epsfbox{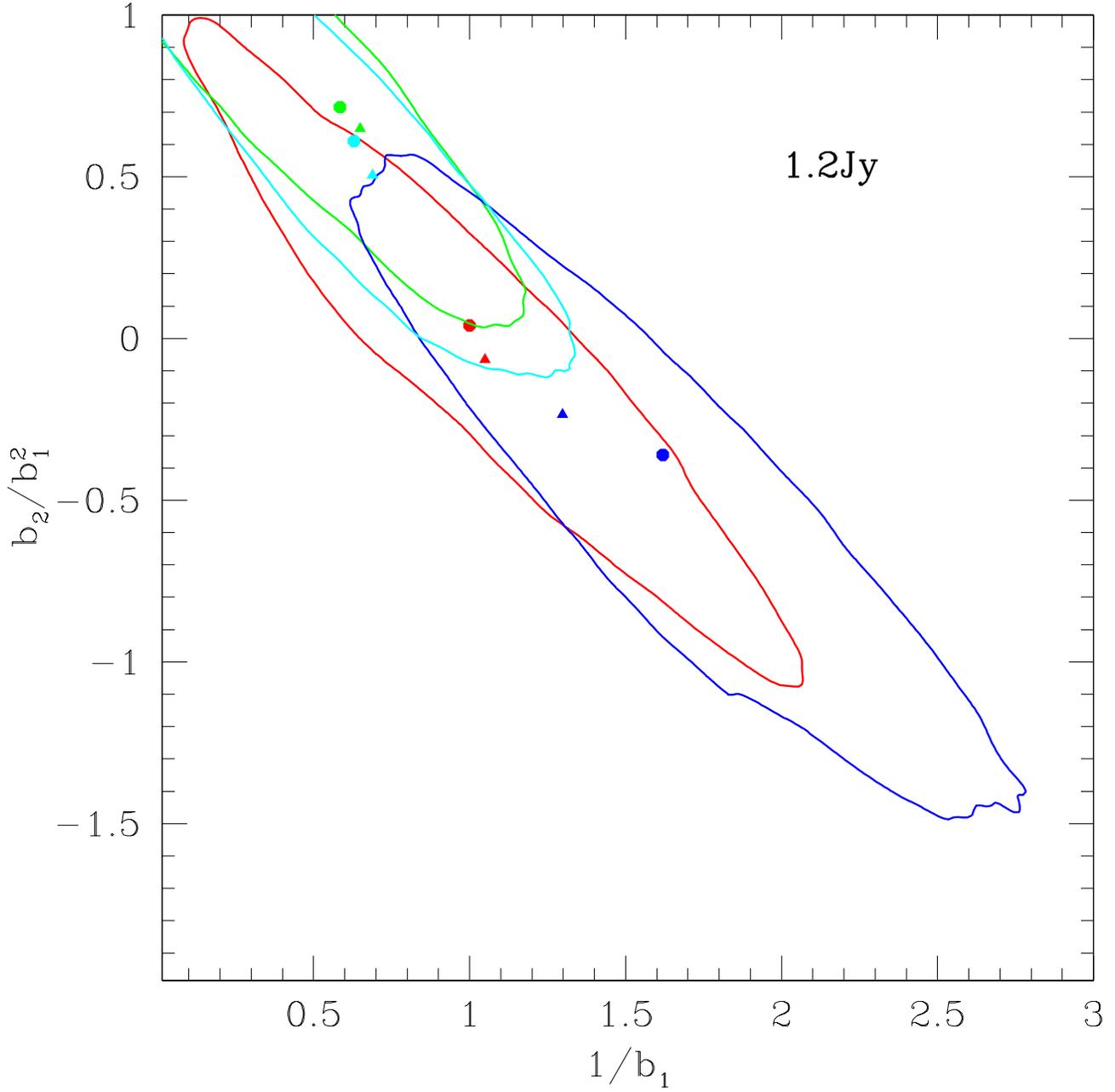}}
\caption{Maximum likelihood estimation from stochastic non-linear bias
realizations of the 1.2Jy survey, corresponding to those shown in
Fig.~\ref{fig_bias}. Contours are $68\%$, triangles denote the
parameters shown in Fig.~\ref{fig_bias}. The maximum likelihood values
(circles) and marginalized error bars are shown in Table~\ref{biast}.}
\label{fig_recov_bias}
\end{figure}

Figure~\ref{fig_recov_bias} shows an example of the use of
Eq.~(\ref{like}) to recover bias parameters. We use parent 2LPT
realizations of $200^3$ particles in a 600 Mpc/h box, which are then
used to select about $2\times 10^6$ ``galaxies'' according to the
prescription in Eq.~(\ref{proba}). These realizations are then used to
generate mock catalogs of the 1.2Jy survey. We generated 200 mock
catalogs for each of the biasing schemes shown in Fig.~\ref{fig_bias},
and then measure $Q$ in redshift space. Three of the models are
generated from $\Lambda$CDM ($\sigma_8=0.7$, $\Omega_m=0.3$,
$\Omega_\Lambda=0.7$) and a fourth model with same power spectrum shape
and normalization but $\Omega_m=1$ (top right panel). The models in the
right panels have about the same $\beta \equiv \Omega_m^{0.6}/b_1
\approx 0.65$. We now describe the results of analyzing these biased
samples with the likelihood in Eq.~(\ref{like}) where all the
quantities ($P_{i}, \bar{Q}_j, \Delta Q_j$) are evaluated from the
{\em unbiased} ($b_1=1$, $b_2=0$) realizations of the surveys
described above to test the assumptions behind our likelihood
analysis.

The plots in Fig.~\ref{fig_recov_bias} show typical outcomes of the
maximum likelihood procedure; triangles denote the ``correct'' biasing
parameters shown in Fig.~\ref{fig_bias} (obtained by measuring the
bispectrum in the parent 2LPT realization). The best fit parameters
and error bars from these particular single realizations of biased
catalogs are shown in Table~\ref{biast}. Note that, as expected, the
derived error bars increase for the antibiased samples. It is also
important to note that although the bispectrum can break the
degeneracy between models with the same $\beta$ but different bias and
$\Omega_m$, for the 1.2Jy survey the errors are large enough that the
two models considered overlap to within $68\%$. These results
are only meant to illustrate the expected error bars and do not
constitute a test of the likelihood analysis (since one can always
find a single realization out of 200 which shows reasonable
reconstruction of the biasing parameters). To test the accuracy of the
likelihood analysis, we checked whether $68\%$ likelihood ellipses
enclosed the correct answer in $68\%$ of the realizations. We found
that in practice this number is about $70\%$; if we instead use a
Gaussian likelihood (including correlations) we found that this number
drops to about $45\%$. Correlations between triangles account for a
significant area of the ellipses in Fig.~\ref{fig_recov_bias},
neglecting correlations leads to error ellipses that contract by a
factor of order 10 in area.

\begin{figure}[t!]
\centering
\centerline{\epsfxsize=18truecm\epsfysize=18truecm\epsfbox{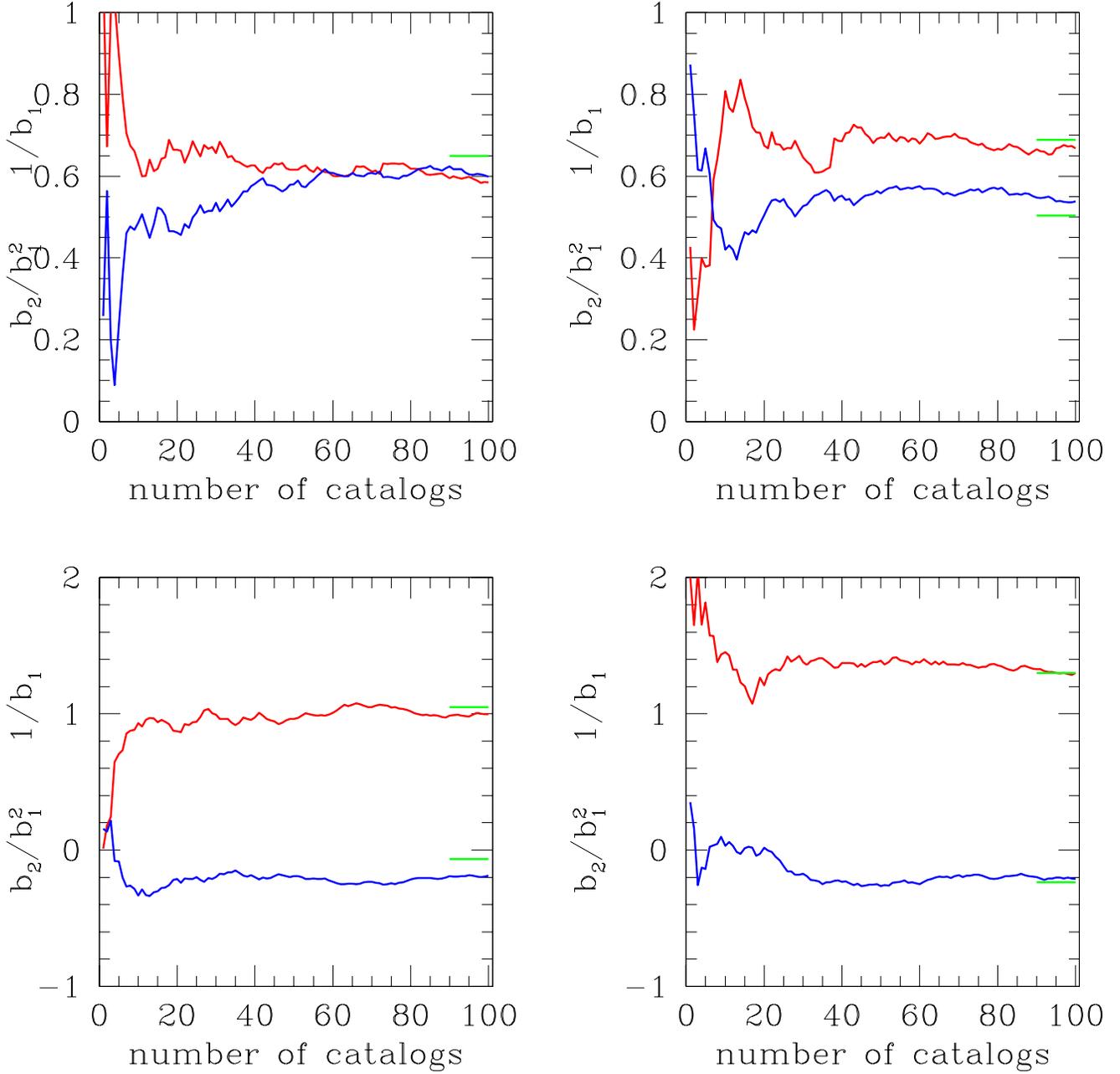}}
\caption{Maximum likelihood estimates of bias parameters as a function
of number of 1.2Jy mock catalogs used. The horizontal marks on the
right denote the values of $b_2/b_1^2$ and $1/b_1$ from a parent 2LPT
realization as shown in Fig.~\protect\ref{fig_bias}.}
\label{fig_ml_bias}
\end{figure}

The use of maximum likelihood (ML) to estimate the best fit bias
parameters is not guaranteed to give statistically unbiased results,
since the likelihood is not Gaussian. To test this,
Fig.~\ref{fig_ml_bias} shows the ML estimates as a function of the
cumulative number of mock catalog realizations used, with panels
ordered as in Fig.~\ref{fig_bias}. The deviations observed at low
numbers of catalogs is due to the effect discussed above, i.e. one
expects individual realizations which are off the underlying value
(error bars for each curve in Fig.~\ref{fig_ml_bias} are avoided for
reasons of clarity, but can be estimated from the ones in
Table~\ref{biast} by scaling by $1/\sqrt{N_{\rm cat}}$). We see that
as the number of catalogs $N_{\rm cat}$ becomes large (and the error
bars become small), the results generally approach the expected
asymptotic value of $1/b_1$, $b_2/b_1^2$ (shown as horizontal marks in
Fig.~\ref{fig_ml_bias}; see also Fig.~\ref{fig_bias} and
Table~\ref{biast}). However, there are detectable (small) estimation
biases, in particular for the non-linear bias parameter. For the lower
left panel, the resulting parameters actually make the agreement in
Fig.~\ref{fig_bias} better. Since these estimation biases are about
$3-4$ times smaller than the expected error bars for a single survey,
we can safely ignore this problem, as it appears not to be systematic.
For larger surveys, where the errors are expected to be smaller it is
also expected that the likelihood will be better approximated by a
Gaussian, thus the estimation bias from ML is also likely to be
smaller, so it is not clear this is a serious problem for constraining
bias parameters, but it is worth keeping in mind.

\subsection{Comparison with Previous Work}

A likelihood analysis for obtaining bias parameters from bispectrum
measurements has been pioneered by Matarrese et al. (1997; MVH), and
extended to redshift-space by Verde et al. (1998). Their approach 
differs from ours in several ways, as they intended their treatment to
next-generation redshift surveys where some of issues faced in this
work will not be as serious (although this will have to be checked in
each particular case).  First, they assume a Gaussian likelihood for
the bispectrum, which as discussed above breaks down in the case of
IRAS surveys. This approximation will considerably improve for surveys
such as 2dF and SDSS.

Second, we consider all triangle shapes with their full covariance
matrix, rather than using only equilateral and colinear triangles to
make the covariance approximately diagonal. In fact, for IRAS surveys
Fig.~14 shows that triangles of similar shapes and scales are strongly
correlated by the window of the survey and sparse sampling. In
addition, discarding triangle shapes throws away information, which is
undesirable. This is of course not a limitation of MVH's method; on
the other hand, analytic calculation of the covariance matrix between
general triangle shapes including the effects of survey geometry and
sampling quickly becomes complicated if one intends to go beyond the
Gaussian approximation. In our case this is handled automatically by
the numerical 2LPT realizations.

Our treatment of perturbation theory (PT) leads to some minor
improvements.  By being Lagrangian, our calculation includes loop
corrections beyond leading order, although only approximately. Their
treatment using second-order Eulerian PT is only consistent for the
bispectrum, not its covariance (which has contributions from the
sixth-point function and thus requires fifth-order Eulerian PT);
however the leading order contribution is Gaussian, so to leading
order (most of the contribution at the scales of interest) both
treatments should agree.  2LPT is not exact beyond second-order in
Eulerian space, however the results of Table~2 show that it gives a
very good approximation to the higher-order moments.

The most important difference is the treatment of redshift
distortions, our numerical calculations allow us to perform the
redshift-space mapping exactly (rather than perturbatively), which
makes quantitative difference even at large scales (e.g. solid versus
dotted lines in Fig.~3). Moreover, by using non-linear dynamics rather
than a phenomenological model to treat non-linear distortions, we
avoid introducing auxiliary quantities such as the velocity dispersion
parameter $\sigma_v$ (Verde et al. 1998; Scoccimarro et al. 1999).

Our calculations of the finite volume effects are in reasonable
agreement with the results found for one-point higher-order statistics
(Szapudi et al. 1999a, 1999b). In particular, we find that the PDF of
the bispectrum is generally non-Gaussian with positive skewness,
whereas Szapudi et al. (1999b) find that the PDF of the skewness
parameter $S_3$ can be approximated by a suitably scaled lognormal. We
also find that in general the estimator bias for $Q$ is smaller than
the error $\Delta Q$, although in the case of the 2Jy survey both
quantities are comparable. It is thus important to correct for
estimator bias (Hui \& Gazta\~naga 1999), even when it is smaller than
the error, since the systematic shift affects the performance of the
likelihood estimation; i.e. without this correction $68\%$ likelihood
contours would not enclose the correct answer $68\%$ of the time. Our
approach can be also applied to one-point statistics, and would be
interesting to compare with the results by Szapudi et
al. (1999a,1999b) since they involve different approximations.

\section{Summary and Conclusions}

We studied the use of the bispectrum to recover bias parameters and
constrain non-Gaussian initial conditions in realistic redshift
surveys. We considered the effects of stochastic non-linear biasing,
radial redshift-distortions, survey geometry and sampling on the shape
dependence of the bispectrum. 

We found that bias stochasticity does not seem to affect the use of
the bispectrum to recover the mean biasing relation between galaxies
and mass, at least for models in which the scatter is uncorrelated at
large scales. Clearly more work is necessary along these lines, we
have just considered a very simple extension to the local
deterministic bias model. We also found that the radial nature of
redshift distortions changes the bispectrum monopole compared to the
plane-parallel values by only a few percent, well below current error
bars.

On the other hand, survey geometry leads to finite volume effects
which must be taken into account in current surveys before comparison
with theoretical predictions can be made.  Similarly, sparse sampling
and survey geometry correlate different triangles leading to a
breakdown of the Gaussian likelihood approximation.  We developed a
likelihood analysis using bispectrum eigenmodes, calculated by Monte
Carlo realizations of mock surveys generated with second-order
Lagrangian perturbation theory (2LPT) and checked against N-body
simulations.  This can be used to derive quantitative constraints from
current and future redshift surveys. In a companion paper (Scoccimarro
et al. 2000) we apply the results obtained here to the analysis of the
bispectrum in \IRAS surveys.

Our treatment can be improved in several ways for applications to
surveys that demand more accuracy. In particular, we relied on the
simple relation in Eq.~(\ref{Qg}) to relate the galaxy bispectrum to
the mass bispectrum; in practice, however, redshift-distortions and
biasing do not conmute, and this relation is more complicated
(Scoccimarro, Couchman \& Frieman 1999).  Related to this issue is a
formulation, so far lacking, of a joint likelihood for the power
spectrum and bispectrum, which involves dependence on $\Omega_m$ and
biasing parameters. In addition, we have only considered closed
triangles in Fourier space. In practice, due to breakdown of
translation invariance by the survey geometry, selection function, and
radial redshift distortions, open triangles also contain useful
information. In order to do this, one possibility is to construct a
general cubic estimator for the bispectrum, analogous to the quadratic
estimator for the power spectrum (e.g. Tegmark et al. 1998), which
seems a rather difficult task for non-Gaussian fields. However, future
large redshift surveys such as 2dF and SDSS will be the ideal datasets
to take full advantage of the cosmological information available from
higher-order statistics, and thus the effort in developing more
accurate statistical methods will likely be rewarded.

\acknowledgments

I thank my collaborators on the \IRAS bispectrum, H.~Feldman,
J.A.~Frieman and J.N.~Fry, for comments and discussions.  I also would
like to thank S.~Colombi, E.~Gazta\~naga and L.~Hui for conversations
on finite volume effects, R.~Crittenden for help with improving the
speed of my bispectrum code, R.~Sheth for discussions about galaxy
biasing, and C.~Murali, S.~Prunet and especially M.~Zaldarriaga for
many helpful discussions regarding maximum likelihood estimation. The
N-body simulations generated for this work were produced using the
Hydra $N$-body code (Couchman, Thomas, \& Pearce 1995). I thank
H.~Couchman and R.~Thacker with help regarding the use of Hydra. This
work was supported by endowment funds from the Institute for Advanced
Study.

\clearpage

\begin{deluxetable}{rcccccc}
\tablewidth{9cm} 
\tablecaption{Ratio $r_p$ of $S_p$ parameters ($3\leq p \leq 8$) in
the ZA to their exact values as a function of spectral index
$n$.\label{ZASp}} 
\tablehead{\colhead{$n$} & \colhead{$r_3$} & \colhead{$r_4$} 
& \colhead{$r_5$} & \colhead{$r_6$} & \colhead{$r_7$} & 
\colhead{$r_8$} }
\startdata
-3  & 0.82  & 0.66  & 0.52  & 0.41  & 0.32  & 0.25  \nl
-2  & 0.77  & 0.58  & 0.42  & 0.30  & 0.22  & 0.15  \nl
-1  & 0.70  & 0.45  & 0.28  & 0.17  & 0.10  & 0.06  \nl
 0  & 0.54  & 0.25  & 0.10  & 0.04  & 0.01  & 0.006  \nl
\enddata
\end{deluxetable}

\begin{deluxetable}{rcccccc}
\tablewidth{9cm} 
\tablecaption{Ratio $r_p$ of $S_p$ parameters ($3\leq p \leq 8$) in
2LPT to their exact values as a function of spectral index
$n$.\label{2LPTSp}}
\tablehead{\colhead{$n$} & \colhead{$r_3$} & \colhead{$r_4$} 
& \colhead{$r_5$} & \colhead{$r_6$} & \colhead{$r_7$} & 
\colhead{$r_8$} }
\startdata
-3  & 1     & 0.98  & 0.95  & 0.92  & 0.89  & 0.86  \nl
-2  & 1     & 0.96  & 0.92  & 0.87  & 0.82  & 0.77  \nl
-1  & 1     & 0.93  & 0.84  & 0.74  & 0.65  & 0.57  \nl
 0  & 1     & 0.80  & 0.54  & 0.35  & 0.24  & 0.17  \nl
\enddata
\end{deluxetable}

\begin{deluxetable}{rcccccc}
\tablewidth{9cm} 
\tablecaption{Ratio $r_p$ of $S_p$ parameters ($3\leq p \leq 8$) in
3LPT to their exact values as a function of spectral index
$n$.\label{3LPTSp}}
\tablehead{\colhead{$n$} & \colhead{$r_3$} & \colhead{$r_4$} 
& \colhead{$r_5$} & \colhead{$r_6$} & \colhead{$r_7$} & 
\colhead{$r_8$} }
\startdata
-3  & 1     & 1     & 0.99  & 0.99  & 0.99  & 0.98  \nl
-2  & 1     & 1     & 0.99  & 0.98  & 0.97  & 0.96  \nl
-1  & 1     & 1     & 0.98  & 0.95  & 0.92  & 0.88  \nl
 0  & 1     & 1     & 0.89  & 0.74  & 0.56  & 0.41  \nl
\enddata
\end{deluxetable}

\begin{deluxetable}{ccc}
\tablewidth{9cm} 
\tablecaption{Percentage of shell-crossing grid points ($x_{sc}$) and
ratios of second to first-order displacements  as a function of power
spectrum cutoff ($k_c$ in h/Mpc) for $\Lambda$CDM ($\sigma_8=0.7$) as in
Fig.~\protect \ref{fig_sc}.\label{sctable}}
\tablehead{\colhead{$k_c$} & \colhead{$x_{sc}$} & 
\colhead{$\left<|{\bf\Psi}_2|\right>/\left<|{\bf\Psi}_1|\right>$} } 
\startdata
0.3 & 1.80 & 0.19 \nl 
0.4 & 7.32 & 0.23 \nl
0.5 & 14.89 & 0.26 \nl
$\infty$ & 42.15 & 0.32 \nl
\enddata
\end{deluxetable}

\begin{deluxetable}{lcccc}
\tablewidth{9cm} 
\tablecaption{Selection function parameters (see
Eq.~(\protect\ref{selfun})) for different \protect\IRAS mock
catalogs. \label{sfpar}} 
\tablehead{\colhead{Catalog} & \colhead{$A$} & \colhead{$r_0$} &
\colhead{$a$} & \colhead{$b$} } 
\startdata 
QDOT 	& 2.42 & 86.01 & 0.52 & 2.98 \nl 
2 Jy. 	& 3.61 & 69.52 & 0.05 & 3.16 \nl 
1.2 Jy. & 7.87 & 72.65 & 0.43 & 3.28 \nl 
PSCz    & 18.83 & 76.71 & 0.81 & 3.1 \nl 
\enddata
\end{deluxetable}

\begin{deluxetable}{cccc}
\tablewidth{12cm} \tablecaption{Recovery of bias parameters from a
single realization of the 1.2Jy survey with stochastic non-linear bias
(likelihood contours shown in Fig.~\protect\ref{fig_recov_bias}) 
\label{biast}}
\tablehead{\colhead{$1/b_1$ (2LPT)} & \colhead{$1/b_1$ (mock)} &
\colhead{$b_2/b_1^2$ (2LPT)} & \colhead{$b_2/b_1^2$ (mock)} }
\startdata 0.65 & $0.59^{+0.42}_{-0.30}$ & 0.65 &
$0.71^{+0.28}_{-0.36}$ \nl 0.69 & $0.63^{+0.46}_{-0.33}$ & 0.50 &
$0.61^{+0.31}_{-0.48}$ \nl 1.05 & $1.00^{+0.61}_{-0.55}$ & -0.07 &
$0.04^{+0.51}_{-0.67}$ \nl 1.30 & $1.62^{+0.69}_{-0.60}$ & -0.24 &
$-0.36^{+0.52}_{-0.79}$ \nl \enddata
\end{deluxetable}

\clearpage

\clearpage

\clearpage

\clearpage

\clearpage

\clearpage

\clearpage

\clearpage

\clearpage

\clearpage

\clearpage

\clearpage

\clearpage

\clearpage

\clearpage

\clearpage

\clearpage

\clearpage

\clearpage

\begin{thebibliography}{}

\bibitem{AMM97} 
Antoniadis I., Mazur P. O., Mottola E.  1997, \prl, 79, 14

\bibitem[Baumgart \& Fry 1991]{BaFr91}
Baumgart D. J., Fry J. N. 1991, \apj,375, 25 

\bibitem[Bernardeau 1992]{Bernardeau92}
Bernardeau F. 1992, \apj, 392, 1

\bibitem[Bernardeau 1994a]{Bernardeau94a}
Bernardeau F. 1994a, \aap, 291, 697

\bibitem[Bernardeau 1994b]{Bernardeau94c}
Bernardeau F. 1994b, \apj, 433, 1

\bibitem[Blanton et al. 1999]{BCOS99}
Blanton M., Cen R., Ostriker J. P., Strauss M. A. 1999, \apj, 522, 590 

\bibitem[Bouchet et al. 1992]{BJCP92}
Bouchet F. R.,  Juszkiewicz R.,  Colombi S., Pellat R. 1992, 
\apjl, 394, L5

\bibitem[Bouchet et al. 1995]{BCHJ95}
Bouchet F. R., Colombi S., Hivon E.,  Juszkiewicz R. 1995, 
\aap, 296, 575

\bibitem[Buchert et al. 1994]{BMW94}
Buchert T.,  Melott A. L., Wei$\beta$ A. G. 1994, 
\aap, 288, 349

\bibitem[Buchalter et al. 2000]{BKJ00}
Buchalter A.,  Kamionkowski M., Jaffe A.  2000, \apj, 530, 36

\bibitem[Catelan \& Moscardini 1994]{CaMo94}
Catelan, P., Moscardini, L. 1994, \apj, 426, 14

\bibitem[Cole et al. 1998]{CHWF98}
Cole S., Hatton S.,  Weinberg D. H.,  Frenk C. S. 1998, 
\mnras, 300, 945

\bibitem[Coles 1993]{Coles 1993}
Coles P. 1993, \mnras, 262, 1065

\bibitem[Colombi et al. 1994]{CBS94}
Colombi S., Bouchet F. R., Schaeffer R. 1994, \aap, 281, 301

\bibitem[Colombi et al. 1995]{CBS95}
Colombi S., Bouchet F. R., Schaeffer R. 1995, \apjs, 96, 401

\bibitem[Colombi et al. 1996]{CBH96}
Colombi S., Bouchet F. R., Hernquist L. 1996, \apj, 465, 14

\bibitem[Colombi et al. 1998]{CSS98}
Colombi S.,  Szapudi I., Szalay A. S. 1998, \mnras, 296, 253

\bibitem[Colombi et al. 1999]{CSJC99} Colombi S., Szapudi I., Jenkins
A., Colberg J. 1999, submitted to MNRAS, {\sf astro-ph/9912236}.

\bibitem[Couchman, Thomas \& Pearce 1995]{CTP95}
Couchman H. M. P., Thomas P. A., Pearce F. R. 1995, 
\apj, 452, 797

\bibitem[Davis \& Peebles 1977]{DaPe77}
Davis, M., \& Peebles, P. J. E. 1977, \apjs, 34, 425

\bibitem[Dekel \& Lahav 1999]{DeLa99}
Dekel, A., \& Lahav, O. 1999, \apj, 520, 24

\bibitem[Eisenstein \& Zaldarriaga 1999]{EiZa99}
Eisenstein, D., \& Zaldarriaga, M. 1999, {\sf astro-ph/9912149}.

\bibitem[Feldman et al. 1994]{FKP94}
Feldman H. A.,  Kaiser N.,  Peacock J. 1994, \apj, 426, 23

\bibitem[Frieman \& Gazta\~naga 1999]{FrGa99}
Frieman J. A.,  Gazta\~naga E. 1999, \apj, 521, L83

\bibitem[Fry \& Seldner 1982]{FrySeldner82}
Fry J. N., Seldner M. 1982, \apj, 259, 474

\bibitem[Fry 1984]{Fry84b}
Fry J. N. 1984, \apj, 279, 499

\bibitem[Fry \& Gazta\~naga 1993]{FrGa93}
Fry J. N., Gazta\~naga E. 1993, \apj, 413, 447

\bibitem[Fry 1994]{Fry94a}
Fry J. N. 1994, \prl, 73, 215

\bibitem[Fry \& Scherrer 1994]{FrSc94}
Fry J. N.,  Scherrer R. J. 1994, \apj, 429, 36

\bibitem[Fry \& Thomas 1999]{FrTh99}
Fry, J. N., \& Thomas, D., 1999, \apj, 524, 591

\bibitem[Fry et al. 1993]{FMS93}
Fry J. N., Melott A. L., Shandarin S. F. 1993, \apj, 412, 504

\bibitem[Gazta\~naga 1994]{Gaztanaga94}
Gazta\~naga E. 1994, \mnras, 268, 913

\bibitem[Goroff et al. 1986]{GGRW86}
Goroff M. H.,  Grinstein B.,  Rey S.-J.,  Wise M. B. 1986, 
\apj, 311, 6

\bibitem[Grinstein \& Wise 1987]{GrWi87}
Grinstein B.,  Wise M. B. 1987, \apj, 320, 448

\bibitem[Groth \& Peebles 1977]{GrPe77}
Groth, E.~J. \& Peebles, P.~J.~E. 1977, \apj, 217, 385

\bibitem[Hamilton \& Culhane 1996]{HaCu96}
Hamilton A. J. S., Culhane M. 1996, \mnras, 278, 73

\bibitem[Hamilton 1997]{Hamilton97a}
Hamilton A. J. S. 1997, \mnras, 289, 285

\bibitem[Hamilton 1998]{Hamilton98} Hamilton A. J. S. 1998, in The
Evolving Universe, ed. D. Hamilton, Kluwer, Dordrecht, p. 185.

\bibitem[Hamilton 2000]{Hamilton00}
Hamilton A. J. S. 2000, \mnras, 312, 257

\bibitem[Hamilton \& Tegmark 2000]{HaTe00}
Hamilton A. J. S., Tegmark M. 2000, \mnras, 312, 285

\bibitem[Heavens \& Taylor 1995]{HeTa95}
Heavens A. F., Taylor A. N. 1995, \mnras, 275, 483

\bibitem[Hivon et al. 1995]{HBCJ95}
Hivon E.,  Bouchet F. R., Colombi S., Juszkiewicz R. 1995, 
\aap, 298, 643

\bibitem[Hui \& Gazta\~naga 1999]{HuGa99}
Hui L., Gazta\~naga E. 1999, \apj, 519, 622

\bibitem[Jing \& B\"orner 1998]{JiBo98}
Jing, Y.P., B\"orner, G. 1998, \apj, 503, 37

\bibitem[Juszkiewicz et al. 1993]{JBC93}
Juszkiewicz R.,  Bouchet F. R., Colombi S. 1993, 
\apjl, 412, L9

\bibitem[Juszkiewicz et al. 1995]{JWACB95} 
Juszkiewicz R.,  Weinberg D. H.,  Amsterdamski P., Chodorowski M., 
Bouchet F. R. 1995, \apj, 442, 39

\bibitem[Kaiser 1987]{Kaiser87}
Kaiser N. 1987, \mnras, 227, 1

\bibitem[KiSt 1998]{KiSt98}
Kim R. S., Strauss M. A. 1998, \apj, 493, 39

\bibitem{KBHP89} Kofman, L., Blumenthal, G. R., Hodges, H., \&
Primack, J. R. 1989, in ``Large-Scale Structures and Peculiar Motions
in the Universe'', eds. Latham, D. W., \& da Costa, L. A. N., 339

\bibitem{LiMu97} Linde, A.D., \& Muhanov, V. 1997, \prd, 56, 535

\bibitem[Mann et al. 1998]{MPH98}
Mann, R. G., Peacock, J. A., \& Heavens, A. F. 1998, \mnras, 293, 209

\bibitem[Matarrese et al. 1997]{MVH97}
Matarrese S.,  Verde L.,  Heavens A. F. 1997, \mnras, 290, 651

\bibitem[Matsubara 1999]{Matsubara99}
Matsubara, T. 1999, \apj, 525, 543

\bibitem[Meiksin \& White 1999]{MeWh99}
Meiksin A., White M. 1999, \mnras, 308, 1179

\bibitem[Moutarde et al. 1991]{MABPR91}
Moutarde F.,  Alimi J.-M.,  Bouchet F. R.,  Pellat R., Ramani A. 1991,  
\apj, 382, 377

\bibitem[Munshi et al. 1999]{MBMS99}
Munshi D., Bernardeau F., Melott A. L., Schaeffer R. 1999, \mnras, 303,
433 

\bibitem[Narayanan et al. 2000]{NBW00}
Narayanan V. K., Berlind A. A., Weinberg D. H. 2000, \apj, 528, 1

\bibitem[Peebles \& Groth 1976]{PG76}Peebles, P. J. E. \& Groth, 
E. J. 1976, A\&A, 53, 131 

\bibitem[Peebles 1980]{Peebles80}
Peebles, P.J.E., 1980, The Large Scale
Structure of the Universe. Princeton University Press, Princeton

\bibitem{Peebles76} Peebles, P.J.E., 1976, \apj, 205, 318

\bibitem{Peebles97} Peebles, P.J.E., 1997, \apj, 483, L1

\bibitem[Scherrer \& Weinberg 1998]{ScWe98}
Scherrer, R. J., Weinberg, D. H. 1998, \apj, 504, 607

\bibitem[Scoccimarro \& Frieman 1996a]{SF95}
Scoccimarro, R., \& Frieman, J. A. 1996a, \apjs, 105, 37

\bibitem[Scoccimarro 1998]{Scoccimarro98}
Scoccimarro R. 1998, \mnras, 299, 1097

\bibitem[Scoccimarro et al. 1998]{SCFFHM98}
Scoccimarro R.,  Colombi S., Fry J. N., Frieman J. A.,  Hivon E., Melott
A. 1998, \apj, 496, 586

\bibitem[Scoccimarro et al. 1999]{SCF99}
Scoccimarro R.,  Couchman H. M. P., Frieman J. A. 1999, \apj, 517, 531

\bibitem[Scoccimarro et al. 1999]{SZL99}
Scoccimarro R., Zaldarriaga M., Hui L. 1999, \apj, 527, 1

\bibitem[Scoccimarro 2000]{Scoccimarro00} 
Scoccimarro R. 2000, submitted to \apj, {\sf astro-ph/0002037}.

\bibitem[Scoccimarro et al. 2000]{SFFF00} Scoccimarro R., Feldman
H. A., Fry J. N., Frieman J. A. 2000, submitted to \apj, {\sf
astro-ph/0004087} (Paper II).

\bibitem[Sheth \& Lemson 1999]{SL99}
Sheth, R. K., \& Lemson, G. 1999, \mnras, 304, 767

\bibitem[Szalay et al. 1998]{SML98}
Szalay A. S., Matsubara T., Landy S. D. 1998, \apj, 498, L1

\bibitem[Szapudi \& Colombi 1996]{SzCo96}
Szapudi I.,  Colombi S. 1996,  \apj, 470, 131

\bibitem[Szapudi et al. 1999]{SCB99}
Szapudi I.,  Colombi S.,  Bernardeau F., 1999, \mnras, 310, 428

\bibitem[Szapudi et al. 1999]{SCJC99}
Szapudi I.,  Colombi S.,  Jenkins A., Colberg J., 1999, submitted to
MNRAS, {\sf astro-ph/9912238}.

\bibitem[Tegmark et al. 1998]{THSVS}
Tegmark M. 2000, Hamilton A. J. S., Strauss, M., Vogeley, M., Szalay,
A. 1998, \apj, 499, 555

\bibitem[Verde et al. 1998]{VHMM98}
Verde L.,  Heavens A. F., Matarrese S.,  Moscardini L.1998, 
\mnras, 300, 747

\bibitem[Zaroubi \& Hoffman 1996]{ZaHo96}
Zaroubi S.,  Hoffman Y. 1996,  \apj, 462, 25

\bibitem[Zel'dovich 1970]{Z70}
Zel'dovich, Ya. B., 1970, A\&A, 5, 84

\end{thebibliography}
\end{document}